%% file: ms.tex
\documentclass[sigconf,nonacm]{acmart}
\pdfoutput=1
\setcopyright{none} % No copyright notice required for submissions

\usepackage{epsfig}
\usepackage{endnotes}
\usepackage{amssymb}
\usepackage{amsmath}
\newtheorem{definition}{Definition}
\usepackage{import}
\usepackage{color}
\usepackage[linesnumbered]{algorithm2e}
\usepackage{pifont}
\newcommand{\cmark}{\ding{51}}

\usepackage{listings}
\usepackage{booktabs}
\usepackage{multirow}
\usepackage[switch]{lineno}
\usepackage{makecell}
\usepackage{subcaption}
\usepackage{tikz-qtree}
\usepackage{pifont}
\usepackage{threeparttable}

\usepackage{caption}
\usetikzlibrary{trees}
\usepackage[normalem]{ulem}

\tikzset{edge from parent/.style=
{draw, edge from parent path={(\tikzparentnode.south)
-- +(0,-4pt)
-| (\tikzchildnode)}},
blank/.style={draw=none},every node/.append style={align=left},every tree node/.style={anchor=north},level distance=0.30in,sibling distance=.06in}

\newcommand{\code}[1]{{\ttfamily #1}}

\newcommand{\nbVirusTotalEngines}{34}
\newcommand{\nbStaticAnalyzers}{36}
\newcommand{\nbDynamicAnalyzers}{5}
\newcommand{\nbAnalyzers}{41} % \nbStaticAnalyzers + \nbDynamicAnalyzers
\newcommand{\nbSamplesSize}{1,395}
\newcommand{\nbBenignsSize}{81}
\usepackage{enumitem}
\setlist[enumerate]{noitemsep,topsep=2pt,leftmargin=*}
\setlist[itemize]{noitemsep,topsep=2pt,leftmargin=*}
\begin{document}

\title{Easy to Fool? Testing the Anti-evasion Capabilities of\\ PDF Malware Scanners}

\author{Saeed Ehteshamifar}
%\authornote{any note?}
%\orcid{i dunno where in the paper it's displayed}
\affiliation{%
  \institution{TU Darmstadt, Germany}
}
\email{salpha.2004@gmail.com}

\author{Antonio Barresi}
\affiliation{%
  \institution{xorlab, Switzerland}
}
\email{antonio@barresi.net}

\author{Thomas R. Gross}
\affiliation{%
  \institution{ETH Zurich, Switzerland}
}
\email{trg@inf.ethz.ch}

\author{Michael Pradel}
\affiliation{%
  \institution{TU Darmstadt, Germany}
}
\email{michael@binaervarianz.de}

\maketitle

\subsection*{Abstract}
\import{./}{abstract.tex}

\import{./}{introduction.tex}
\import{./}{methodology.tex}

\import{./}{implementation.tex}
\import{./}{analyzers.tex}

\import{./}{results.tex}

\import{./}{discussion.tex}

\import{./}{related.tex}

\import{./}{conclusion.tex}

\section*{Acknowledgment}
This work was supported by the German Federal Ministry of Education and
Research and by the Hessian Ministry of Science and the Arts within
CRISP, by the German Research Foundation within the ConcSys and Perf4JS
projects, and by the Hessian LOEWE initiative within the
Software-Factory 4.0 project.
The authors would also like to thank InfoGuard and the malware scanner vendors that anonymously participated in this study. Moreover, we thank Mathias Payer and other anonymous reviewers for their reviews and feedback.

\bibliography{ms}

\newpage
\appendix
\import{./}{academic_analyzers.tex}

\end{document}

%% file: abstract.tex
Malware scanners try to protect users from opening malicious documents by statically or dynamically analyzing documents.
However, malware developers may apply evasions that conceal the maliciousness of a document.
Given the variety of existing evasions, systematically assessing the impact of evasions on malware scanners remains an open challenge.
This paper presents a novel methodology for testing the capability of malware scanners to cope with evasions.
We apply the methodology to malicious Portable Document Format (PDF) documents and present an in-depth study of how current PDF evasions affect \nbAnalyzers{} state-of-the-art malware scanners.
The study is based on a framework for creating malicious PDF documents that use one or more evasions.
Based on such documents, we measure how effective different evasions are at concealing the maliciousness of a document.
We find that many static and dynamic scanners can be easily fooled by relatively simple evasions and that the effectiveness of different evasions varies drastically.
Our work not only is a call to arms for improving current malware scanners, but by providing a large-scale corpus of malicious PDF documents with evasions, we directly support the development of improved tools to detect document-based malware.
Moreover, our methodology paves the way for a quantitative evaluation of evasions in other kinds of malware.

%% file: introduction.tex
\section{Introduction}

Malware scanners, or shortly scanners, are software tools that detect malicious files, or in brief, malware.
Two common types of scanners are static and dynamic scanners.
\emph{Static scanners} reason about a file by examining its content without actually running it.
In contrast, \emph{dynamic scanners} examine the behavior of a file at run-time, either by executing it (e.g. Windows executable), or by opening it in the appropriate application (e.g. Adobe Reader for PDF files) or an emulator of such an application.

Perhaps as old as the emergence of scanners~\cite{historyofevasion} are \emph{evasions}, which are used by attackers to circumvent scanners.
Also known as ``logic bombs'' in earlier work~\cite{greenberg1998mobile}, evasions try to fool scanners through a variety of static techniques, such as code obfuscation, and dynamic techniques, such as checking the run-time environment to behave benignly when the environment appears to be a scanner.
The ultimate goal is the same across all evasions: bypass the scanner, while preserving the infection capabilities of the file to compromise the victim's security.

As scanners are constantly improving their abilities to detect malware, evasion techniques are evolving as well. To bypass modern defenses that deploy both static and dynamic analysis, attackers may combine evasions, which can lead to side-effects that have to be assessed.
Vendors of malware scanners must keep fighting new evasion techniques and their combinations, just like new attacks.
It is therefore crucial for vendors to understand which evasions to address first and how evasions and their combinations impact their scanners.

In this work we present a systematic methodology to quantitatively study and compare evasions.
The methodology is applicable to any type of malware and their corresponding scanners.
The main goal of the methodology is to determine how effective evasions are, or to put inversely, how effective scanners are despite the presence of evasions.
In addition, the methodology allows for measuring unintended side-effects of an evasion, e.g., turning an undetected file into a detected one, and for measuring the effect of combining multiple evasions.

We use our methodology to study evasions for PDF files.
Document-based malware attacks are a prevailing problem~\cite{pdf_cve_statistics, officeonrise, exploit_CVE_2018_4990}. These attacks use email or web traffic to deliver malicious documents to victim systems. Then they compromise the system's security by exploiting a vulnerability in the document processing application (e.g., a PDF exploit) or by using legitimate features of the document processing application itself (e.g., embedding an executable file). The attacker's goal is to execute arbitrary machine code or code in a powerful language supported by the client applications (e.g., Visual Basic scripts for Office files). As most organizations need to be able to receive or download files in different document formats, these attacks are particularly difficult to prevent compared to attacks that use executable file formats only. Yet, malicious documents are as powerful as malicious executables because they can lead to arbitrary code execution.

Unfortunately, despite the widespread use of document files and works that study their evasions~\cite{carmony2016extract, Maiorca2013, xu2016automatically, Dang2017, zhang2016adversarial, biggio2013evasion, laskov2014practical},
little is currently known about the effectiveness of document evasion techniques, their combinations, and the dependence of evasion effectiveness on other malware components, such as the exploit used by a malicious document.
Using our methodology, we study evasion techniques for PDFs 
and evaluate their effectiveness in bypassing state-of-the-art PDF scanners.
To this end, we develop a novel framework, called Chameleon, that enriches existing malicious 
PDF documents with one or more evasions.
Chameleon automatically creates PDF exploits with evasions and validates whether the generated exploits work successfully despite the evasion.
Based on \nbSamplesSize{} documents generated by Chameleon, we study 
\nbAnalyzers{} widely used PDF scanners (\nbVirusTotalEngines{} of which are available via VirusTotal) and report a detailed analysis of the results.

The findings of our study include the following:
\begin{itemize}
  \item
    Except for one studied scanner~\cite{2018arXiv181012490J}, none of the \nbAnalyzers{} scanners is immune to evasions. 
    Each of them can be fooled by some evasions into misclassifying a 
    malicious document as benign.
    This result is particularly surprising because the vulnerabilities exploited in our malicious documents have been known for several years.

  \item
    There are huge variations across different scanners.
    While some scanners identify most malicious documents despite evasions, 
    other scanners are fooled by more than 80\% of all evasions.

  \item
    We identify three combinations of evasions that are particularly 
    dangerous as they can mislead all but two scanners.

   \item
     The attack mechanism used in a document influences the effectiveness of 
     evasions.
     For example, an exploit that relies only on JavaScript can often be 
     effectively concealed by obfuscating the JavaScript code.

   \item
     Evasions can be easily combined
     in an automated way to bypass both static and dynamic scanners.

   \item
     Evasions may have side effects and can become counterproductive by 
     making scanners suddenly detect an otherwise undetected malicious document.
\end{itemize}

The results of this study are relevant for several groups of people.
First, our methodology will help researchers to study and rank evasions by their effectiveness in a consistent manner.
Moreover, our study sheds light on the anti-evasion problems that state-of-the-art document scanners suffer from.
Second, vendors of security scanners, e.g., anti-virus or sandbox solution vendors, can learn and use our findings to further harden their solutions against evasion techniques.
Third, users and organizations that need to defend themselves against malware attacks obtain a better understanding of how effective their deployed security solutions are, particularly for PDF-based attacks.
We believe that publicly sharing the knowledge about evasions and their effectiveness is the best step toward effectively mitigating potential attacks.
In addition, we are closely collaborating with vendors of scanners to make them aware of their current weaknesses.

In summary, we make the following contributions:
\begin{itemize}
    \item \textbf{Evasion assessment methodology:} We propose a methodology to quantitatively study the effectiveness of evasions on a large scale. This methodology can be used for all types of malware and their corresponding scanners.
    \item \textbf{Chameleon framework:} We implement our methodology for PDF exploits in Chameleon, a novel framework that automatically transforms malicious PDF documents into evasive documents.
    \item \textbf{A benchmark test suite:} We make a corpus of \nbSamplesSize{} evasive PDF files generated by Chameleon publicly available, to foster future work on evaluating and improving PDF security scanners.
    \item \textbf{An in-depth study of evasions for document-based malware:} We conduct a large-scale study of the effectiveness of 19 PDF evasions on a set of~\nbAnalyzers{} scanners. Our findings show widely used scanners to be easily fooled by evasions, motivating work on better-coping with evasions.
    
\end{itemize}

%% file: methodology.tex
\section{A Methodology for Assessing Evasions}
\label{ss:methodology}

To study the anti-evasion capabilities of malware scanners, we present a generic methodology to study evasions and their effect on scanners.
The methodology is designed to address a set of research questions presented in Section~\ref{ss: questions}.
To address these questions, we define several metrics that measure how evasions influence the outcome of malware scanners (Section~\ref{ss: metrics}).
Our methodology assumes that possibly malicious files are analyzed by scanners, and that these files may contain evasions.
\emph{File} here means any type of file ranging from executables (e.g., Android apps) to document files (e.g., Office documents).
\emph{Scanner} here means a software tool that classifies a file either as malicious or as benign.
Finally, \emph{evasion} refers to a technique aimed at concealing the fact that a file is malicious.

\subsection{Research Questions}
\label{ss: questions}

We focus on the following research questions (RQs):

\begin{itemize}
    \item \textbf{RQ1:} How accurately do the scanners classify malicious and benign files in the presence of evasions?
    \item \textbf{RQ2:} How effective are the evasions at fooling specific scanners?
    \item \textbf{RQ3:} Which evasions are most effective?
    \item \textbf{RQ4:} Do some evasions have the opposite of the expected effect, i.e.,
      do they cause scanners to detect  malicious files that are missed otherwise?
    \item \textbf{RQ5:} Are there combinations of evasions that are harder to detect than the individual evasions?
    \item \textbf{RQ6:} Does the effectiveness of an evasion depend on the exploit or the payload used in a malicious file?
\end{itemize}

\subsection{Metrics for Assessing Evasions}
\label{ss: metrics}

To address the above questions, we define several metrics.
For illustration, consider the set of example files in Table~\ref{t:example suite}.
There are two single evasions, called $e_1$ and $e_2$, and one evasion that is a combination of the two, called $e_{1,2}$.
Lines 1--16 in the table represent malicious files.
Each malicious file is based on an exploit, a payload, and optionally, also an evasion.
The last two lines of the table represent two benign files without any payload, exploit, or evasion.
For each file, the table shows if two (hypothetical) scanners classify the file as malicious or benign.

\begin{table}[tb]
\centering
\caption{Example files to illustrate the metrics.}
\setlength{\tabcolsep}{7pt}
\footnotesize
\begin{tabular}{rlllll}
\toprule
\# & Evasion & Exploit & Payload & \multicolumn{2}{c}{Scanner outcome} \\
\cmidrule{5-6}
&&&& $s_1$ & $s_2$ \\
\midrule
1   &   --      &   $x_1$   &   $p_1$   &   malicious   & benign    \\
2   &   --      &   $x_1$   &   $p_2$   &   malicious   & benign    \\
3   &   --      &   $x_2$   &   $p_1$   &   malicious   & malicious    \\
4   &   --      &   $x_2$   &   $p_2$   &   malicious   & malicious    \\

5   &   $e_{1}$      &   $x_1$   &   $p_1$   &   malicious   & malicious    \\
6   &   $e_{1}$      &   $x_1$   &   $p_2$   &   malicious   & malicious    \\
7   &   $e_{1}$      &   $x_2$   &   $p_1$   &   malicious   & benign    \\
8   &   $e_{1}$      &   $x_2$   &   $p_2$   &   malicious   & benign    \\

9   &   $e_{2}$      &   $x_1$   &   $p_1$   &   benign   & benign    \\
10   &   $e_{2}$      &   $x_1$   &   $p_2$   &   benign   & benign    \\
11   &   $e_{2}$      &   $x_2$   &   $p_1$   &   malicious   & benign    \\
12   &   $e_{2}$      &   $x_2$   &   $p_2$   &   malicious   & benign    \\

13   &   $e_{1,2}$      &   $x_1$   &   $p_1$   &   benign   & malicious    \\
14   &   $e_{1,2}$      &   $x_1$   &   $p_2$   &   benign   & malicious    \\
15   &   $e_{1,2}$      &   $x_2$   &   $p_1$   &   benign   & malicious    \\
16   &   $e_{1,2}$      &   $x_2$   &   $p_2$   &   benign   & malicious    \\

17   &   --      &   --   &   --   &   benign   & benign    \\
18   &   --      &   --   &   --   &   malicious   & benign    \\

\bottomrule
\end{tabular}
\label{t:example suite}
\end{table}

For a  set $E$ of evasions, a set $S$ of scanners, and a set $F$ of malicious files, 
we use the notation $f^e$ for a file $f \in F$ that uses an evasion $e \in E$.
The function $\mathit{mal}: S \times F \rightarrow \mathit{Boolean}$  indicates whether a scanner classifies a given file as malicious.
Inversely, the notation $\neg \mathit{mal}(s,f)$ means that a scanner $s$ classifies a file $f$ as benign.
For a set $F_{ben}$ of benign files that do not contain any payload, exploit, or evasion, and a set $F$ of malicious files from a set of payloads,
optionally a set of exploits,
and a set $E$ of evasions,
we define the following formulas to measure scanners' performance in dealing with evasions.

\textbf{Recall and false positive ratio.}
We evaluate a scanner's accuracy in distinguishing malicious from benign files by measuring recall and false positive (FP) ratio.

\begin{definition}[Recall]
Given a scanner $s$ and a set $F$ of malicious files, the recall of $s$ is:
$$
\mathit{recall}(s, F) = \frac{|\{ f \in F ~|~ \mathit{mal}(s,f) \}|}{|F|}
$$
\label{def: recall}
\end{definition}
The FP ratio is computed for each scanner $s$ on a set $F_{ben}$ of benign files.
Here we use benign files only, because these are the files where a scanner might mislead users by erroneously classifying a file as malicious.
\begin{definition}[FP ratio]
Given a scanner $s$ and a set $F_{ben}$ of benign files, the FP ratio of $s$ is:
$$
\mathit{FP~ratio}(s, F_{ben}) = \frac{|\{ f \in F_{ben} ~|~ \mathit{mal}(s,f) \}|}{|F_{ben}|}
$$
\label{def: fp ratio}
\end{definition}

For the example files in Table~\ref{t:example suite}, recall and FP ratio are as follows.

\vspace{.5em}
{
\centering
$recall(s_1) = \frac{10}{16} = 0.625$ \hspace{2em} $recall(s_2) = \frac{8}{16} = 0.5$ \\
}
\vspace{.5em}

\vspace{.5em}
{
\centering
$FP~ratio(s_1) = \frac{1}{2} = 0.5$ \hspace{2em} $FP~ratio(s_2) = \frac{0}{2} = 0.0$\\
}
\vspace{.5em}

\textbf{Effectiveness of evasions.} We measure the success of an evasion in bypassing a scanner for a set of malicious files.
Intuitively, we compare the outcome of a scanner for each file with an evasion to the outcome of the scanner for the exact same file without the evasion.

\begin{definition}[Evasion effectiveness]
Given a scanner $s$ and a set $F$ of malicious files, the effectiveness of an evasion $e$ is:
$$
\mathit{eff}(e, s, F) = \frac
{|\{ f \in F ~|~ \mathit{mal}(s,f) \wedge \neg \mathit{mal}(s,f^{e}) \}|}
{|\{ f \in F ~|~ \mathit{mal}(s,f) \}|} \\
$$
If the scanner does not classify any file as malicious, which would lead to a division by zero, the effectiveness is defined to be zero.
\label{def: eff}
\end{definition}

For example, to compute the effectiveness of $e_{1}$ in Table~\ref{t:example suite}, we
compare row 5 with row 1, row 6 with row 2, row 7 with row 3, and row 8 with row 4.
As a result, the effectiveness for the two scanners $s_1$ and $s_2$ is:

\vspace{.5em}
{
	\centering
	$\mathit{eff}(e_{1}, s_1, F) = \frac{0}{4} = 0.0$
	\hspace{2em} 
	$\mathit{eff}(e_{1}, s_2, F) = \frac{2}{2} = 1.0$\\
}
\vspace{.5em}

We summarize the effectiveness of evasions across multiple scanners
by computing the arithmetic mean of the effectiveness values across these scanners.

For the running example in Table~\ref{t:example suite}, the evasion effectiveness for scanners $S=\{s_1,s_2\}$ is calculated as follows:
\begin{gather*}
\mathit{eff}(e_{1}, \{s_1, s_2\}, F) = \frac{0.0 + 1.0}{2} = 0.5 \\
\mathit{eff}(e_{2}, \{s_1, s_2\}, F) = \frac{\frac{2}{4} + \frac{2}{2}}{2} = 0.75 \\
\mathit{eff}(e_{1,2}, \{s_1, s_2\}, F) = \frac{\frac{4}{4} + \frac{0}{2}}{2} = 0.5
\end{gather*}

Likewise, to summarize the effectiveness across a set of evasions, we average effectiveness values across the set. For the evasions $E=\{e_1,e_2,e_{1,2}\}$ in the example we have:
\begin{gather*}
\mathit{eff}(\{e_{1}, e_{2}, e_{1,2}\}, \{s_1, s_2\}, F) = \frac{0.5 + 0.75 + 0.5}{3} \approx 0.58
\end{gather*}

\textbf{Counter-effectiveness: attacker's cost of using evasions.} Even though the goal of an evasion is to bypass the detection of a scanner, an evasion may also have the opposite effect: to cause a scanner mark a
file as malicious that otherwise would be marked as benign.
We call an evasion that has such an effect \emph{counter-effective}.

\begin{definition}[Evasion counter-effectiveness]
\label{def:counterEff}
Given a scanner $s$ and a set $F$ of malicious files, the counter-effectiveness of an evasion $e$ is:
\begin{multline*}
\mathit{counterEff}(e, s, F) =
\frac
{|\{ f \in F ~|~ \neg \mathit{mal}(s, f) \wedge \mathit{mal}(s, f^e) \}|}
{|\{ f \in F ~|~ \neg \mathit{mal}(s, f) \}|}
\end{multline*}
\end{definition}

For example, for the evasions in Table~\ref{t:example suite} we have:
\begin{gather*}
\mathit{counterEff}(e_1, \{s_1, s_2\}, F) = \frac{\frac{0}{0} + \frac{2}{2}}{2} = 0.5 \\
\mathit{counterEff}(e_2, \{s_1, s_2\}, F) = \frac{\frac{0}{0} + \frac{0}{2}}{2} = 0.0 \\
\mathit{counterEff}(e_{1,2}, \{s_1, s_2\}, F) = \frac{\frac{0}{0} + \frac{2}{2}}{2} = 0.5
\end{gather*}

From the attacker's perspective, another cost of using evasions is that they may interfere with the malicious behavior of a file.
For example, an evasion in a PDF document that requires the user to move the mouse before the malicious behavior is triggered may not only hide the maliciousness but also reduce the chance that the attack is successful.
One way of measuring this cost would be to conduct a user study that measures how often a file with an evasion successfully triggers the attack when the file is handled by users.
We leave the challenge of measuring this cost for future work.

\textbf{Added effectiveness by combining evasions.} A set $E$ of evasions that is combined in a file adds to the effectiveness of the individual evasions in $E$ only if $E$ is effective but none of the subsets of $E$ are effective.
We formalize this idea in the following metric.

\begin{definition}[Evasion added effectiveness]
Given a scanner $s$ and a set $F$ of malicious files, the added effectiveness of a combined evasion $e$ is:
\begin{multline*}
\mathit{addedEff}(e, s, F) =\\
\frac
{|\{ f \in F ~|~ \mathit{mal}(s,f) \wedge \neg \mathit{mal}(s,f^e) \wedge \nexists e' \subset e ~.~ \neg \mathit{mal}(s,f^{e'}) \}|}
{|\{ f \in F ~|~ \mathit{mal}(s,f) \wedge \neg \mathit{mal}(s,f^e) \}|}
\end{multline*}
where $e' \subset e$ refers to the single or combined evasions that constitute $e$.
\label{def: added eff}
\end{definition}

For example in Table~\ref{t:example suite}, the added effectiveness of $e_{1,2}$, which is composed of $e_1$ and $e_2$, is:
\begin{gather*}
\mathit{addedEff}(e_{1,2}, s_1, F) = \frac{2}{4} = 0.5 \\
\mathit{addedEff}(e_{1,2}, s_2, F) = \frac{0}{0} = 0.0
\end{gather*}

\textbf{Dependence of evasions on other components of a malware.}
To study how the effectiveness depends on other malware components, e.g., its payload or exploit,
Definition~\ref{def: eff} can be applied to subsets of all files that have that particular component in common.
For example, to study how the effectiveness depends on the exploit used in a malicious PDF file, effectiveness is computed on the subset of all PDF files under study that are based on that very exploit.
In Table~\ref{t:example suite}, to focus on those files that use exploit $x_2$, we consider only the set of files $F_{x_2}$ in rows 3, 4, 7, 8, 11, 12, 15, and 16.
The effectiveness of evasion $e_1$ w.r.t.\ these files across all scanners is:
\begin{gather*}
\mathit{eff}(e_{1}, \{s_1, s_2\}, F_{x_2}) = \frac{\frac{0}{2} + \frac{2}{2}}{2} = 0.5
\end{gather*}

\medskip
The set of metrics defined above allows us to address the research questions given in Section~\ref{ss: questions}.
In particular, Definitions~\ref{def: recall} and~\ref{def: fp ratio} address RQ1, Definition~\ref{def: eff} addresses RQ2--3 and RQ6, Definition~\ref{def:counterEff} addresses RQ4, and Definition~\ref{def: added eff} addresses RQ5.

%% file: implementation.tex
\section{Chameleon Framework: Generating Malicious, Evasive PDFs}
\label{ss: implementation}

Our study is based on various evasions, which we summarize in a taxonomy (Section~\ref{ss: evasions}).
To systematically study how these evasions evade malware scanners, we present Chameleon, a framework to automatically create malicious files that contain one or more evasions.
Our implementation of the framework focuses on malicious PDF documents.
Such malware is particularly interesting because document-based malware attacks are a prevailing problem~\cite{pdf_cve_statistics, officeonrise, exploit_CVE_2018_4990} and because the ability of PDF scanners to cope with evasions is currently understudied.

\begin{figure}
    \centering
    \includegraphics[width=0.5\textwidth]{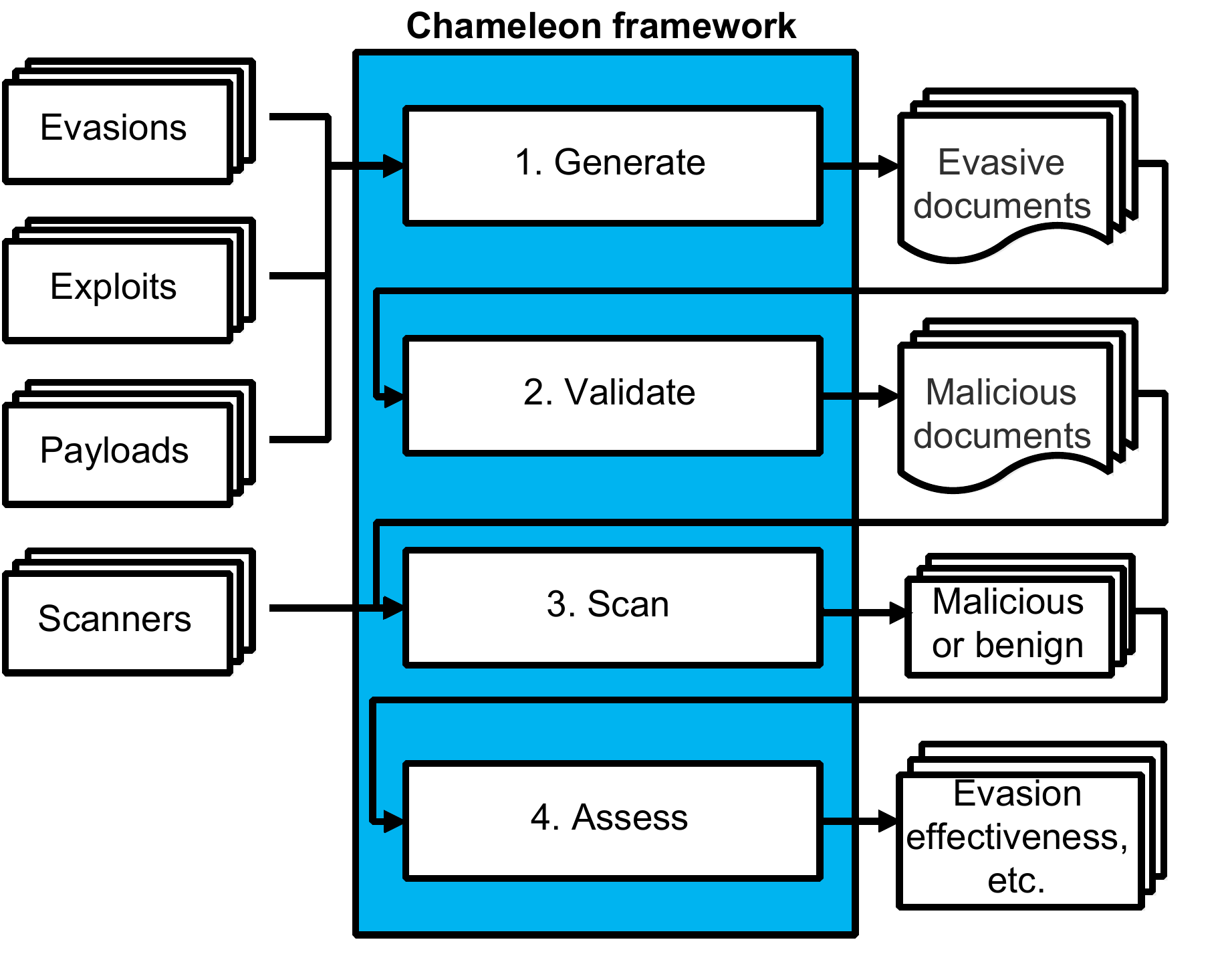}
    \caption{Overview of the Chameleon framework and its four steps.}
    \label{fig: chameleon overview}
\end{figure}

Figure~\ref{fig: chameleon overview} shows an overview of the Chameleon framework.
The inputs to the framework are a set of evasions, a set of \emph{exploits}, i.e., code 
that uses a bug or vulnerability, and a set of \emph{payloads}, i.e., code that contains the malicious behavior of the attack.
We discuss these inputs in Sections~\ref{ss: evasions}, \ref{ss: exploits}, and~\ref{ss: payloads}, respectively.
Given these inputs, Chameleon generates evasive PDF documents and validates that they still behave maliciously despite the evasion(s).
We then pass these documents to a set of PDF scanners (Section~\ref{ss: analyzers}) and measure their ability to handle the evasions (Section~\ref{ss: results}).

\subsection{Evasions}
\label{ss: evasions}

Various evasion techniques, for executables and other potentially malicious
file formats have been proposed~\cite{corona2014lux0r,pdfstaticevasion,pdf_obfus_odd_algs,related_static_obfus,vmray_anti_evasion,joe_anti_evasion,lastline_anti_evasion,fireeye_anti_evasion}.
To provide some background on different kinds of evasions and the scope of this 
work, we present a taxonomy of evasions.
The taxonomy tries to cover the major classes of evasions that are relevant for 
malicious documents without claiming to be complete.
In particular, we focus on evasions implemented in high-level languages that 
can be embedded into document formats, such as JavaScript and Visual Basic.

\begin{figure*}[tb]
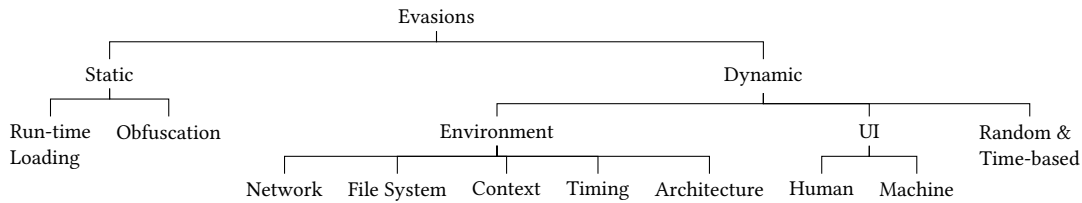

\small
\centering
\Tree 	[.{Evasions}
          [.Static
    				{Run-time \\ Loading}
    				Obfuscation
    			]
    			[.Dynamic 
    				[.Environment
    					Network {File System} Context Timing Architecture
    				]
            [.{UI} Human Machine ]
    				{Random \& \\ Time-based}
          ] ]
\captionsetup{justification=centering}
\caption{A taxonomy of evasion techniques.}
\label{f:taxonomy}
\end{figure*}

Figure~\ref{f:taxonomy} shows an overview of the taxonomy.
We distinguish between dynamic and static evasions.
\emph{Static evasions} attempt to modify the document or the code embedded 
into it in a way that influences a static analysis of the document.
In contrast, \emph{dynamic evasions} change the run-time behavior of a 
document to influence the outcome of a dynamic analysis of the document.

\subsubsection{Static Evasions}
Among the static evasions, there are two broad classes.
First, \emph{run-time loading} tries to conceal malicious behavior by loading 
parts of the code at run-time, making it harder for a static scanner to 
detect the maliciousness.
For example, an evasion based on run-time loading may download the malicious 
payload once the document is opened on the victim's machine, i.e., after 
having been checked by a static scanner.
Second, \emph{obfuscation} modifies the malicious source code to conceal its 
purpose.
There are various obfuscation techniques, such as encryption, multi-pass 
encoding, logical xor, and changing the code 
structure~\cite{pdfstaticevasion}.

\subsubsection{Dynamic Evasions}
Dynamic evasions can be broadly classified into three categories.
The first category are \emph{environment-based evasions}, which attempt to 
take the execution environment into account.
This approach is specially appealing for targeted attacks, where the 
attacker has some information about the target system. We further classify environment-based evasions 
into the following five categories:

\textbf{Network}.
    Dynamic scanners may restrict the network access of documents to prevent malware from downloading its payload.
    Network-based evasions check the network connection to identify the 
    presence of a dynamic scanner or a sandbox.
  
\textbf{File system}.
    Since many dynamic scanners rely on known libraries or executables, the 
    presence of particular files in the file system may disclose a dynamic 
    scanner.
    File system-based evasions check whether particular files exist to 
    decide whether to perform any malicious behavior.

\textbf{Context}.
    Information about the system language, locales, most recently used 
    documents, the time zone, etc.\ can be abused by attackers to target 
    particular victim systems~\cite{le2017broad, rasthofer2017making}.
    A context-based evasion deceives dynamic scanners by behaving 
    maliciously only in particular contexts.

\textbf{Timing}.
    Due to the virtualized environment used by most dynamic scanners, some 
    operations have observably lower performance than other operations.
    For example, the performance difference of a CPU-intensive computation
    and a GPU-intensive computation is higher in a virtual machine than on a 
    physical machine.
    The reason is that in a modern virtual machine, many CPU instructions 
    run natively, whereas translating GPU instructions to physical 
    instructions imposes a noticeable overhead~\cite{ho2014tick}.
    Timing-based evasions exploit such differences in execution time to 
    determine the presence of a dynamic scanner~\cite{timing_evasion}.
    
\textbf{Architecture}.
  These  evasions recognize architectural idiosyncrasies of the 
    underlying physical or virtual machine.
    Examples include an incorrectly return value of the \code{CPUID} instruction in
    QEMU~\cite{ferrie2007attacks} and GPU fingerprinting~\cite{gpufingerprinting}.

\medskip
The second category of dynamic evasions are \emph{UI-based evasions}.
Such evasions monitor interactions with UI elements to determine whether a 
human or a machine is using the system.
We further classify UI-based evasions into two sub-classes:

\textbf{Human user}.
    These evasions attempt to identify a human user and expose the 
    malicious behavior only to such users.
    For example, an evasion may wait until the user scrolls to a 
    particular page or clicks a particular UI 
    element~\cite{userinteraction}.

\textbf{Machine user}.
    Instead of trying to detect a human user, an evasion may also check for 
    evidences that a machine is interacting with the system.
    For example, text entered into a form with a superhuman 
    typing speed or clicks on an invisible element suggests the presence of a machine 
    user~\cite{keragala2016detecting}.

\medskip
The third category of dynamic evasions are \emph{random-based and time-based 
evasions}.
This kind of evasion triggers an attack either probabilistically or 
depending on the current time, e.g., only on specific times of the 
day~\cite{timebased}.

\subsubsection{Implementation of Evasions}
\label{ss: combination of evasions}

\begin{table*}[t]
\caption{Static and dynamic evasions implemented in the Chameleon framework. The last column denotes whether the evasion is implemented in the PDF structure, the embedded JavaScript code, or both.}
\label{t:evasions}
\footnotesize
\begin{tabular}{@{}p{6em}p{6em}p{39em}p{7.5em}@{}}
\toprule
Class & Name & Description & Implementation\\
\midrule
\multicolumn{3}{l}{\emph{Static evasions:}} \\
\midrule

Run-time loading & steganography & Encode the JavaScript code into 
                          an image file embedded in the PDF document. Load 
                          and \code{eval} the code at run-time. & PDF \& JavaScript \\

                          & content & Store the JavaScript code as the content in the 
                        PDF document. Load and \code{eval} the code at run-time. & PDF \& JavaScript \\
\\
JavaScript \mbox{obfuscation} & rev & Lexically reverse the JavaScript code. & JavaScript  \\

                          & xor & Encode the JavaScript code by applying 
                        the bitwise xor operator with the specified key. & JavaScript \\
\\                        
PDF \mbox{obfuscation}    & objstm & Compress the malicious PDF as an Object 
                          Stream and put it into a benign PDF document. & PDF 
                          \\

                          & nest & Recursively embed the malicious PDF into 
                          a benign PDF document for one or more times. & PDF                       
                          \\
                        & decoy & Insert the malicious JavaScript code 
                        into a benign PDF document. In contrast  to ``nest'', this evasion does not recursively nest documents into each other. & PDF
                        \\

\midrule 
\multicolumn{3}{l}{\emph{Dynamic evasions:}}\\
\midrule 
Context               & lang & Trigger if the language of the PDF viewer 
is in the specified set of languages. & JavaScript \\

                          & resol & Trigger if the desktop resolution is 
                          in the specified range. & JavaScript \\
                          
                          & mons & Trigger if the user's computer has the specified 
                          number of monitors attached to it. & JavaScript  \\

                          & filename & Trigger if the generated exploit's 
                          filename has not changed. Some scanners change the filename before the analysis. & JavaScript \\

\\
UI                        & scroll & Trigger when the user has scrolled 
                            to the specified page. & PDF \\

                            & captcha & Trigger if the user's text 
                            input matches the specified string. & JavaScript \\
                            
                            & alert\_three & Show an alert dialog box with 
                            three buttons and trigger if the specified                       
                            button is clicked. & JavaScript \\
                
                            & doc\_close & Trigger when the document gets 
                            closed. & PDF  \\
                            
                            & alert\_one & Show an alert dialog box with one                                
button and trigger when the button is clicked. & JavaScript \\
                            
                            & mouse & Trigger if the mouse position 
                            changes. & JavaScript \\
\\
Random and time-based   & delay & Delay the exploitation for the                                
given amount of time (time bomb). & JavaScript \\

                            & tod & Trigger at the specified time of the day. & JavaScript  
                            \\

\bottomrule
\end{tabular}
\end{table*}

Based on our taxonomy, we have implemented 19 evasions (7 static and 12 dynamic), as summarized in Table~\ref{t:evasions}.
Some evasions take an argument to configure different variants of the evasion.
For example, the ``lang'' evasion can be configured by passing the language to check for, and the ``delay'' evasion can be configured with a specific amount of time.
Using the ``lang'' evasion with ``English'' as the argument will result in a document that attacks only computers with the English version of Adobe Reader:
{\tt \small
\begin{verbatim}
if (app.language == "English")
  exploit(); // trigger the exploit
\end{verbatim}
}

In addition to injecting individual evasions into documents, Chameleon 
also allows to  blend multiple evasions into \emph{combined evasions}.
We refer to combined evasions that contain at least one static and at least 
one dynamic evasion as \emph{hybrid evasions}.
When combining evasions of the same kind, we focus on evasions from different classes, e.g., run-time loading with JavaScript obfuscation.
For UI-based evasions, we also combine several evasions from the same class to gradually increase the complexity of the UI interactions required to trigger the attack.
Moreover, Chameleon creates an evasion that combines several context-based evasions, to create a document that targets a very specific environment and remains silent otherwise.

\subsection{Exploits}
\label{ss: exploits}

Chameleon uses two PDF exploit modules provided by the Metasploit framework\footnote{\url{https://github.com/rapid7/metasploit-framework}} and adapts them to introduce evasions.
The ``Toolbutton'' exploit\footnote{CVE-2013-3346} abuses a use-after-free vulnerability in the implementation of the Adobe-specific JavaScript function \code{app.add\-Tool\-Button}.
The exploit executes some JavaScript code to set up the environment and then triggers the vulnerability by calling the vulnerable function.
To implement dynamic evasions, we trigger the vulnerability only if the condition checked by the evasion holds.

The ``Cooltype'' exploit\footnote{CVE-2010-2883} uses a malicious font file in addition to malicious JavaScript code.
The font file is loaded after the JavaScript code has set up the environment for exploitation.
We slightly modify the exploit by adding an exploitation trigger that controls whether and when the exploit is executed.
The dynamic evasions call this trigger only if the condition checked by evasion holds.

In addition to ``Toolbutton'' and ``Cooltype'', Metasploit provides other PDF exploit modules.
We choose these two exploits as they target a popular PDF reader software (Adobe Reader) and because
they are old and well-known. If our evasions can fool PDF scanners using old and well-studied
exploits then the evasions are at least as or even more effective when applied to more recent
or zero-day exploits.

\subsection{Payloads}
\label{ss: payloads}

Another important component of any attack is the payload that it carries.
As payloads, Chameleon uses native machine code that is executed after the vulnerability is triggered.
We use three different payloads, two provided by Metasploit and one that we develop ourselves.
The first payload, ``Reverse Bind'', establishes a TCP connection to a remote host allowing the remote host to control the exploited machine.
The second payload, ``Powershell'', spawns an instance of Windows Powershell with a command that creates a text file in a temporary directory.
The third payload, ``Exit'', simply exits the Adobe Reader process.

\subsection{Generating and Validating Evasive Documents}

We implement the Generate step of Chameleon on top of the Metasploit framework, which we use to generate exploit documents, and the Origami PDF transformation library\footnote{\url{https://github.com/gdelugre/origami}}, which we use to manipulate documents.
The 19 evasions are implemented as a new Metasploit module, which can be used with any of the PDF exploit modules.

After generating a supposedly malicious document, Chameleon checks that the document is indeed malicious (step Validate).
To this end, Chameleon opens the document in the vulnerable version of Adobe Reader inside a sandbox, interacts with it according to the evasions (e.g., by moving the mouse or waiting for some time), and checks whether the payload is executed.
At the moment this process is mostly but not fully automated because for context-based evasions, the sandbox needs to be manually adapted to the context that an evasion is looking for (e.g., for the ``mons'' evasion, the number of displays attached to the sandbox has to be properly set).

Our implementation and a set of \nbSamplesSize{} generated PDF documents are publicly available.\footnote{\url{https://github.com/sola-da/Chameleon/}}

%% file: analyzers.tex
\section{PDF Scanners}
\label{ss: analyzers}

\begin{table}[tb]
	\caption{PDF scanners used for the study.}
	\label{tab:analyzers}
	\footnotesize
	\setlength{\tabcolsep}{6pt}
    \renewcommand*{\arraystretch}{.6}
	\begin{tabular}{@{}lcccc@{}}
		\toprule
		Scanner & Static & Dynamic & Academic & Commercial \\
		\midrule
		ALYac & \cmark & & & \cmark \\
		AVG & \cmark & & & \cmark \\
		AVware & \cmark & & & \cmark \\
		Ad-Aware & \cmark & & & \cmark \\
		AhnLab-V3 & \cmark & & & \cmark \\
		Antiy-AVL & \cmark & & & \cmark \\
		Arcabit & \cmark & & & \cmark \\
		Avast & \cmark & & & \cmark \\
		Avira & \cmark & & & \cmark \\
		Baidu & \cmark & & & \cmark \\
		BitDefender & \cmark & & & \cmark \\
		CAT-QuickHeal & \cmark & & & \cmark \\
		Cuckoo & & \cmark & \cmark & \cmark \\
		Cyren & \cmark & & & \cmark \\
		DS1 & & \cmark & & \cmark \\% VMRay Analyzer
		DS2 & & \cmark & & \cmark \\% Lastline Analyst
		Emsisoft & \cmark & & & \cmark \\
		F-Prot & \cmark & & & \cmark \\
		F-Secure & \cmark & & & \cmark \\
		Fortinet & \cmark & & & \cmark \\
		GData & \cmark & & & \cmark \\
		Ikarus & \cmark & & & \cmark \\
		Jiangmin & \cmark & & & \cmark \\
		Kaspersky & \cmark & & & \cmark \\
		MAX & \cmark & & & \cmark \\
		McAfee-GW-Edition & \cmark & & & \cmark \\
		MicroWorld-eScan & \cmark & & & \cmark \\
		Microsoft & \cmark & & & \cmark \\
		NANO-Antivirus & \cmark & & & \cmark \\
		PDF-Scrutinizer~\cite{schmitt2012pdf} & & \cmark & \cmark & \\
		Qihoo-360 & \cmark & & & \cmark \\
		Rising & \cmark & & & \cmark \\
		SAFE-PDF~\cite{2018arXiv181012490J} & \cmark & & \cmark & \\
		Slayer~\cite{maiorca2012pattern} & \cmark & & \cmark & \\
		Sophos & \cmark & & & \cmark \\
		SploitGuard & & \cmark & & \cmark \\
		Symantec & \cmark & & & \cmark \\
		Tencent & \cmark & & & \cmark \\
		TrendMicro & \cmark & & & \cmark \\
		VIPRE & \cmark & & & \cmark \\
		ZoneAlarm & \cmark & & & \cmark \\
		\bottomrule
	\end{tabular}
\end{table}

We study \nbStaticAnalyzers{} static and \nbDynamicAnalyzers{} dynamic scanners, including both academic and widely used commercial tools, as listed in Table~\ref{tab:analyzers}.
To categorize a scanner as static or dynamic we rely on information provided 
by the vendors or developers.
Based on this information, we consider a scanner as static if it reasons 
about a PDF document without opening the document in a PDF viewer.
In contrast, dynamic scanners open a PDF document in a PDF viewer or an emulator and then 
analyze its runtime behavior, e.g., by tracking how the PDF viewer interacts 
with the operating system.
Our study includes more static than dynamic scanners because static scanners are more common in practice.

To run the commercial static scanners on our PDF documents, we use the application programming interface (API) of VirusTotal that runs close to 60 static scanners at once on a given document. We ignore those scanners that do not detect any of the exploits we use (perhaps because they are not designed to detect PDF malware), which leaves \nbVirusTotalEngines{} commercial static scanners.
To run the commercial dynamic scanners, we use the individual APIs provided by the respective vendors of these scanners.
The vendors of two commercial dynamic scanners requested to participate anonymously, so we refer to them as DS1 and DS2.
Appendix~\ref{sec:scanner setup} explains the detailed setup of the non-commercial scanners.
In addition to the scanners listed in Table~\ref{tab:analyzers}, we considered several others,
including PDFRate~\cite{smutz2012malicious} and PJScan~\cite{laskov2011static}, but were unable to use them for our study because they either were unavailable or had some issues in our local setup (see Appendix~\ref{sec:other analyzers} for details).

%% file: results.tex
\section{Results}
\label{ss: results}

In this section, we address the research questions from Section~\ref{ss: questions} 
by applying the methodology from Section~\ref{ss: metrics}.
We apply \nbAnalyzers{} widely used PDF scanners to a total of \nbSamplesSize{} malicious PDF documents generated by the Chameleon framework and an additional \nbBenignsSize{} benign PDF documents.
The benign documents comprise train tickets, governmental documents, manuals, tutorials, and some suspicious looking PDF files that are known to be benign.
All documents used for our study will be made available as a benchmark for future work.

\subsection{RQ1: Recall and False Positives}
\label{ss: recall fps}

\begin{figure*}[t]
	\begin{minipage}{0.55\textwidth}
		\hspace{-.6em}\includegraphics{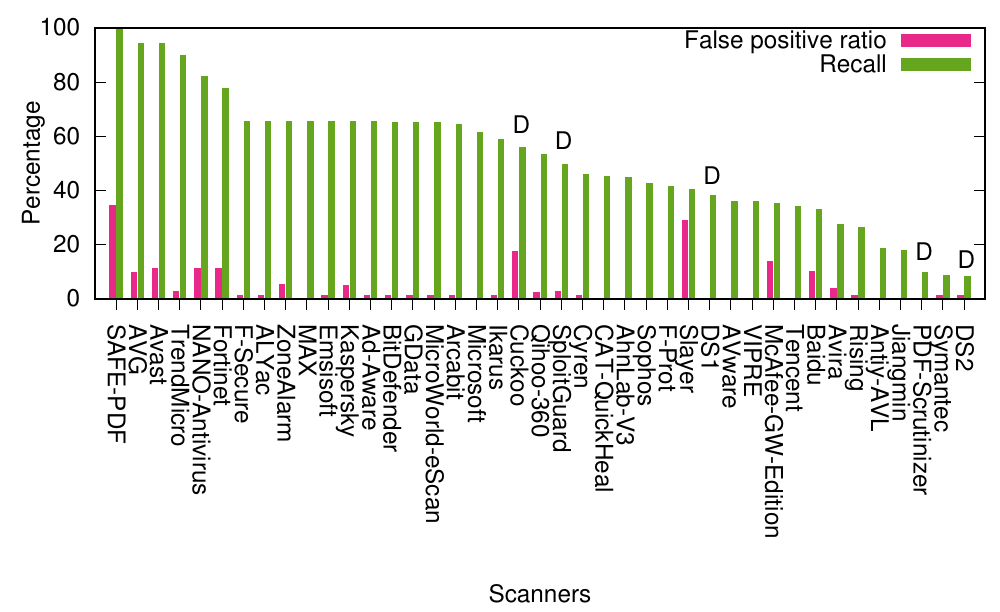}
		\captionof{figure}{Recall and false positive ratio of the studied scanners.  
			The dynamic scanners are marked with ``D''.}
		\label{fig: recall fp ratio}
	\end{minipage}
	\hfill
	\begin{minipage}{0.4\textwidth}
		\hspace{-.6em}\includegraphics{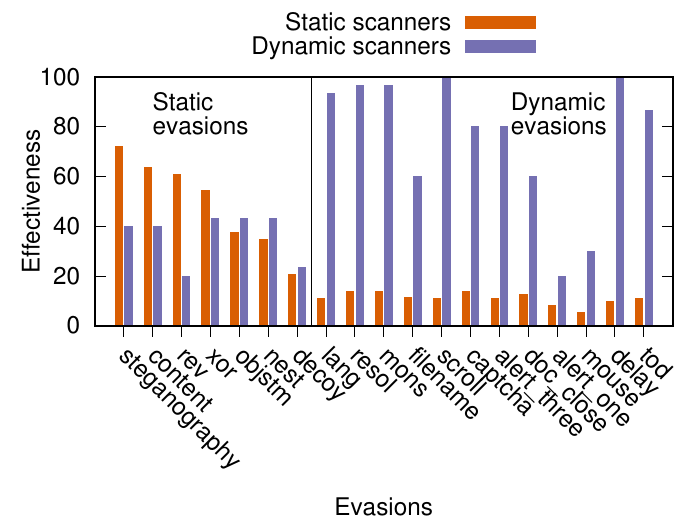}
		\captionof{figure}{Effectiveness for different classes of static and 
			dynamic evasions. The results are averaged over all static 
			(red) and dynamic (blue) scanners.}
		\label{fig: best case eff}
	\end{minipage}		
\end{figure*}

The following addresses RQ1, i.e., how accurately the scanners classify 
documents into malicious and benign in the presence of evasions.
We measure the recall and the false positive 
ratio of each scanner, as described in Section~\ref{ss: metrics}.
Figure~\ref{fig: recall fp ratio} shows the results.
A higher recall means that the scanner is more successful in 
identifying malicious PDF documents, despite the presence of evasions.
The figure shows that almost all studied scanners are affected by evasions, 
as their recall values are lower than 100\%.
Furthermore, we find a huge variation across the studied scanners: While 
some scanners, e.g., SAFE-PDF and AVG, identify all or most malicious documents despite 
evasions, others miss many malicious documents.
Some scanners have a recall lower than 20\%, showing that they are easily 
bypassed by evasions.

In principle, a scanner could achieve 100\% recall by labeling each document 
as malicious.
To address this potential problem, Figure~\ref{fig: recall fp ratio} also 
shows the false positive ratio of each scanner.
We find that all scanners have a false positive ratio below 15\%, except Cuckoo (17.5\%), Slayer (28.77\%), and SAFE-PDF (34.57\%).
There is no strong correlation (Pearson correlation coefficient: 36.61\%) between recall and false positive ratio.
We conclude from these results that none of the scanners tries to boost its 
recall at the cost of precision, which seems reasonable as users easily drop 
a tool if they are overwhelmed with spurious warnings.

\subsection{RQ2: Evasion Effectiveness by Scanner}

\begin{figure*}[tb]
	\begin{minipage}{0.55\textwidth}
		 \begin{subfigure}{\textwidth}
			  	\hspace{-.6em}\includegraphics{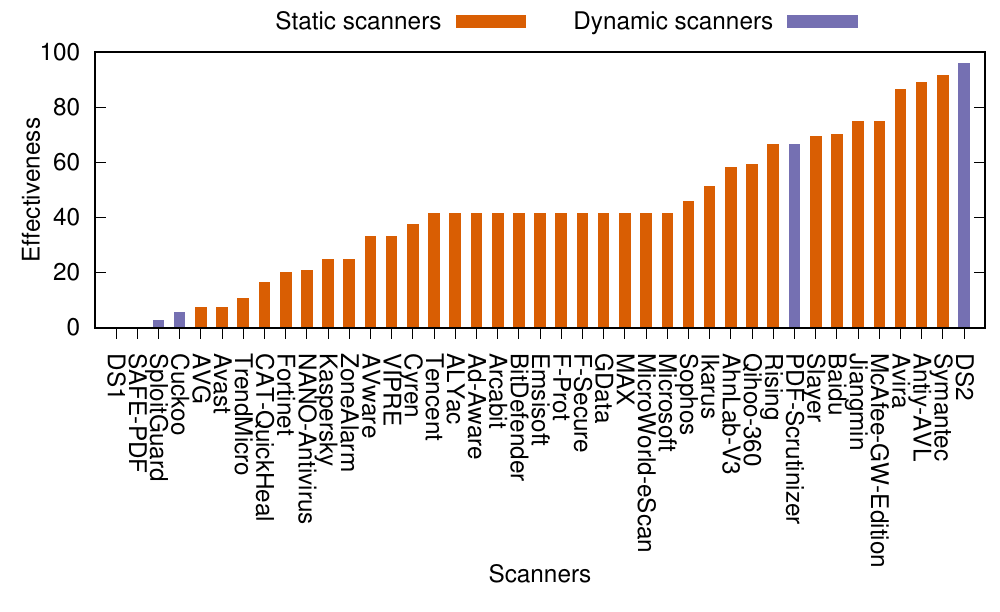}
			  	\caption{Static evasions}
			  	\label{fig: static evasions proj}
		 \end{subfigure}		 
		 \begin{subfigure}{\textwidth}
			  	\hspace{-.6em}\includegraphics{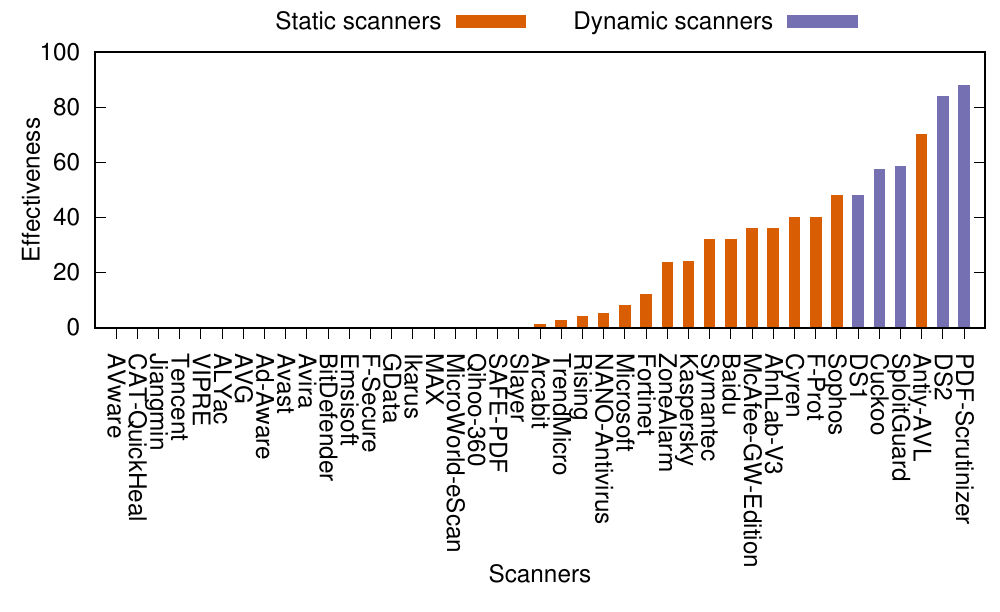}
			  	\caption{Dynamic evasions}
			  	\label{fig: dynamic evasions proj}
		 \end{subfigure}
		 \caption{Per-scanner effectiveness of static and dynamic evasions.}		
	\end{minipage}
	\hspace{1em}
	\begin{minipage}{0.4\linewidth}
		\begin{subfigure}[b]{\textwidth}
			\hspace*{-.6em}\includegraphics{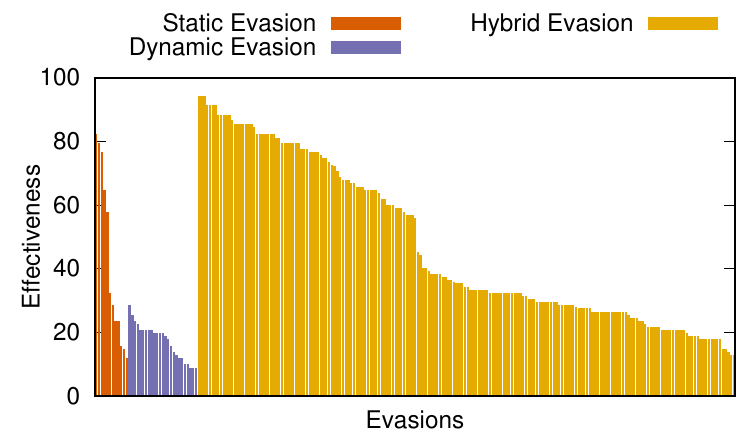}
			\caption{Documents with ``Toolbutton'' exploit.}
			\label{figure:toolbutton projection}
		\end{subfigure}
		\vspace{2em}
		
		\begin{subfigure}[b]{\textwidth}
			\hspace*{-.6em}\includegraphics{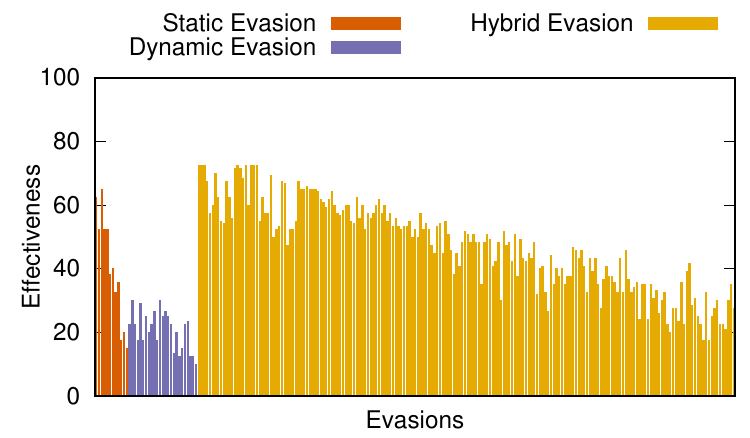}
			\caption{Documents with ``Cooltype'' exploit.}
			\label{figure:cooltype projection}
		\end{subfigure}
		\caption{Effectiveness of evasions for the subsets of malicious documents that use a specific exploit. Each bar corresponds to an evasion with a specific argument. The hybrid evasions, and some of the static and dynamic evasions  result from combining evasions (Section~\ref{ss: combination of evasions}).}
	\end{minipage}
\end{figure*}

To better understand the susceptibility of the scanners to static and dynamic evasions, we assess the effectiveness of evasions in bypassing specific 
scanners (RQ2).
We compute the evasion effectiveness for each 
scanner by averaging the effectiveness across all evasions.
Figures~\ref{fig: static evasions proj} and~\ref{fig: dynamic evasions proj} 
present the results for static and dynamic evasions, respectively.
A lower value indicates that a scanner is less susceptible to evasions.
The results for the static evasions in Figure~\ref{fig: static evasions 
proj} show some interesting effects.
Somewhat surprisingly, the effectiveness of 12 of the~\nbVirusTotalEngines{} VirusTotal scanners, roughly 
in the middle of the figure, is exactly the same, out of which 8 have the exact same recall, too (Figure~\ref{fig: recall fp ratio}).
A possible explanation is that multiple scanners rely on 
the same underlying decision mechanism, e.g., because one scanner queries 
another scanner as part of its decision, or because the same analysis 
algorithm is provided under several brands.
Previous, informal reports claim that some static scanners share their results~\cite{static_analyzers_share_analysis_result}, which our results 
confirm.

Another interesting observation is that a dynamic, not a static, scanner (DS2) is the most susceptible to static evasions.
A comparison of  Figures~\ref{fig: static evasions proj} and~\ref{fig: dynamic evasions proj} shows that DS2 is highly susceptible to both dynamic and static evasions.
This finding suggests that DS2 not only uses dynamic analysis, but also 
heavily relies on static analysis techniques.

The static scanners in the right part of Figure~\ref{fig: dynamic evasions proj} are
impacted by dynamic evasions.
A likely reason is that adding an evasion changes the signature of the PDF documents, and that the scanners rely on these signatures.

\subsection{RQ3: Most Effective Evasions}
\label{ss:best-case effectiveness}

Understanding which evasions are most effective (RQ3) is important both for attackers and for developers of security scanners.
We measure the effectiveness of evasions for each scanner and then compute 
the average over all static and the average over all dynamic 
scanners.
Some evasions take arguments, e.g., the language used by the ``lang'' evasion or the xor key used by the ``xor'' evasion.
For such evasions, we try a range of arguments and report the highest observed effectiveness.

Figure~\ref{fig: best case eff} shows the results.
We sort the evasions as in Table~\ref{t:evasions}.
Overall, the results show that static scanners are much more susceptible to 
static evasions, while dynamic scanners get fooled by dynamic evasions, 
which is unsurprising and confirms our classification of evasions.

Among the static evasions, those related to run-time loading and JavaScript obfuscation are the 
most effective, suggesting that many static scanners rely on checking the 
JavaScript code embedded into PDFs to identify malicious 
behavior.
The high effectiveness of many of the static evasions is somewhat 
surprising given that some of these static evasions have been known 
for several years~\cite{related_static_obfus, corona2014lux0r, more_related_static_obfus}.

For dynamic evasions, attackers can choose from a relatively large set of 
highly effective evasions, including the two time-related evasions, most of 
the context-related evasions, and some of the user interaction-related 
evasions, e.g., ``scroll''.
In contrast, some other UI-related evasions are only moderately effective, e.g., ``mouse''.
The reason is that some of the dynamic scanners use anti-evasion techniques, such as user interaction emulation, to counter these evasions.
For example, Cuckoo moves the mouse after opening a document in a PDF
reader to counter the ``mouse'' evasion~\cite{cuckoo_mouse_movement}.

\begin{table}[]
    \small
    \centering
	\caption{The evasions that bypass all but two scanners. The evasions are applied to a document in the given order.}
	\label{t:detected by two}
    \renewcommand*{\arraystretch}{1.2}
    \setlength{\tabcolsep}{5pt}
    \begin{tabular}{rp{25em}}
    \toprule
    &Combined evasions \\
    \midrule
     1 & mouse, scroll, content, steganography, xor (key: 40) \\
     2 & alert\_one, scroll, content, steganography, xor (key: 40) \\
     3 & alert\_three (trigger on No or Cancel), mouse, mons ($>$1), filename, lang (German), tod (8 AM -- 4 PM), scroll, content, steganography, xor (key: 40) \\
    \bottomrule
    \end{tabular}
\end{table}

To better understand highly effective evasions with their particular arguments across all scanners,
Table~\ref{t:detected by two} lists the evasions that bypass all but two scanners (NANO-Antivirus and SAFE-PDF) for at least one exploit.
All the evasions presented in Table~\ref{t:detected by two} are hybrid, showing that
combinations of static and dynamic evasions are effective as they bypass both static and dynamic scanners at the same time.
Overall, the observation that three different combinations of evasions can circumvent almost all scanners is alarming and motivates future work on anti-evasion techniques.

\subsection{RQ4: Counter-effective Evasions}
\label{ss: Counter-effective Evasions}

In the following we address RQ4, i.e., whether some evasions have the opposite of the expected effect by causing a scanner to detect an otherwise missed malicious document.
To answer the question, we measure the counter-effectiveness of each evasion for each scanner and then compute 
the average over all static scanners and the average over all dynamic 
scanners.
For evasions that take parameters, we report the highest counter-effectiveness observed across a range of values.

\begin{figure}[tb]
    \hspace{-.6em}\includegraphics{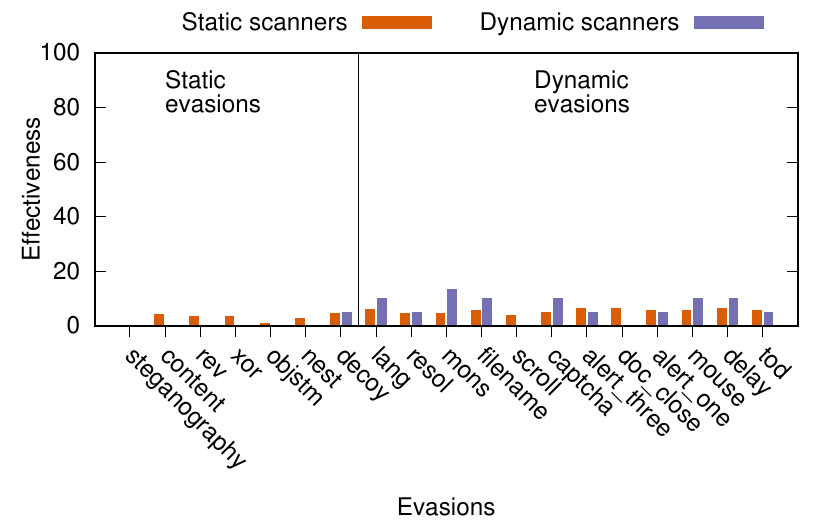}
    \caption{Counter-effectiveness for different classes of static and 
    dynamic evasions. The results are averaged over all static
    (red) and dynamic (blue) scanners.}
    \label{fig: worst case counter-eff}
\end{figure}

Figure~\ref{fig: worst case counter-eff} shows the counter-effectiveness of the studied evasions, sorted as in Figure~\ref{fig: best case eff}.
Perhaps surprisingly, most evasions are, at least sometimes, counter-effective.
A likely reason is that some scanners consider the changes of the document caused by the evasions as indicators of malicious intent.
For example, the context-related evasions add some code to the document to check the
current context, and  some scanners may consider this activity to be suspicious.
In fact, all context-related and time-related evasions are counter-effective.
Furthermore, all evasions are counter-effective for at least some static scanners, with the exception of the ``steganography'' evasion.
This evasion's high effectiveness (Figure~\ref{fig: best
  case eff}) and low counter-effectiveness  should concern the developers of scanners.

Although most evasions are sometimes counter-effective, the counter-effectiveness in Figure~\ref{fig: worst case counter-eff} is relatively low compared to the effectiveness of evasions.
Moreover, we observe counter-effectiveness only in a relatively small subset of the studied scanners.
For static scanners, the evasions behave counter-effectively only on McAfee-GW-Edition, Qihoo-360, and Rising.
For dynamic scanners, all counter-effective behavior that we observe is due to DS2.

\subsection{RQ5: Combinations of Evasions}
\label{ss:combinations}

\begin{figure}[tb]
    \hspace{-.4em}\includegraphics{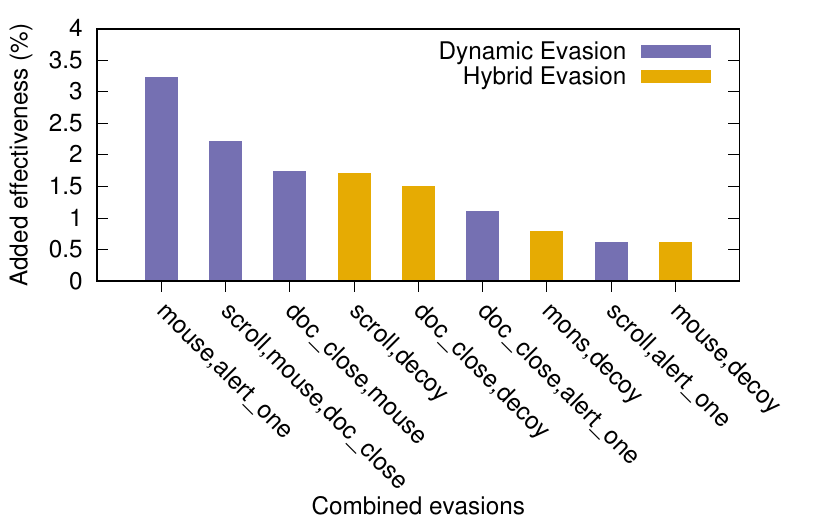}
    \caption{Evasions that have greater than 0.5\% added effectiveness.}
    \label{fig: added eff}
\end{figure}

A combination of evasions may be more effective than the individual evasions.
For example, even though some evasions may not be able to bypass a scanner alone, their combination may be able to do so (RQ5).
In the following we discuss the  added effectiveness of
combined dynamic, combined static, and hybrid evasions.
Figure~\ref{fig: added eff} presents the results for combined evasions with more than 0.5\% added effectiveness.
Averaged over all scanners, the added effectiveness of even the most successful combined evasions is relatively low (about 3.2\%).
For some individual scanners, though, we find higher added effectiveness values.
That is, an attacker interested in bypassing a particular scanner could combine evasions suitable for this task.

Interestingly, combining multiple static evasions does not cause any added effectiveness, suggesting that a single static evasion is sufficient to fool scanners susceptible to this kind of evasion.
Furthermore, all combined dynamic evasions in Figure~\ref{fig: added eff} result from combining UI-based evasions, showing that an evasion that requires a more complicated user interaction is more successful.

\subsection{RQ6: Influence of Exploits and Payloads on Evasion Effectiveness}
\label{ss: influence of exploit effectiveness}

The effectiveness of an evasion may depend on the specific exploit or payload used in a malicious document.
For example, consider an exploit that relies on malicious JavaScript and therefore may be detected by scanners that check the JavaScript code in a document.
For such an exploit, a JavaScript-based evasion may work particularly well, because the evasion reduces the chance that scanners identify the document as malicious.
The following studies to what extent the effectiveness of an evasion depends on the exploit or payload used in the malicious document (RQ6).
To this end, we compute the effectiveness of each evasion for the subset of all documents that use a particular exploit or payload.

\subsubsection{Influence of Exploit}

Figure~\ref{figure:toolbutton projection} shows the evasion effectiveness for documents with the ``Toolbutton'' exploit.
The evasions related to JavaScript obfuscation work particularly well, 
since this exploit is based on malicious JavaScript code only, i.e., no other objects, such as fonts or images, are needed.
Many of these evasions are greater than 80\% effective.
The sudden drop in the effectiveness of static and hybrid evasions is also due to the drastically higher success of JavaScript obfuscation-based versus PDF obfuscation-based evasions.

The evasion effectiveness for documents based on the ``Cooltype'' exploit is shown in Figure~\ref{figure:cooltype projection}.
We use the same order of evasions as in Figure~\ref{figure:toolbutton projection} to enable a comparison between the two exploits.
The results show several interesting effects.
First, the most effective evasions for the ``Toolbutton'' exploit reach almost 100\% effectiveness, whereas the peak effectiveness of ``Cooltype''-based documents is only around 75\%.
The reason is that ``Cooltype'' requires both malicious JavaScript code and a malicious font file to be embedded in the PDF.
As a result, none of the static evasions alone is highly effective at hiding ``Cooltype''-based documents.
Second, the effectiveness of evasions based on PDF obfuscation are higher for ``Cooltype'' than for ``Toolbutton'' (the second half of static evasions' bars in the figure).
This result suggests that many scanners identify the exploit by searching for the
malicious font file and thereby make those evasions effective that change the signature of
the PDF document (and hence the signature of the embedded font file).

\subsubsection{Influence of Payload}

With the same approach as exploits, we study the dependence of the evasions on the payload.
We compute the effectiveness of the subset of the samples with each of the three payloads.
In contrast to the exploits, we do not observe any major differences in effectiveness of the evasions.

\medskip
\noindent
Overall, studying the influence of exploits and payloads on the effectiveness of evasions shows that exploits and evasions may influence each other.
Developers of PDF scanners should be aware of this influence when developing anti-evasion techniques, as an attacker might choose suitable evasions depending on how a PDF exploit works.

%% file: discussion.tex
\section{Discussion}
\label{ss: discussion}

In this section, we discuss how security scanners can defend against evasions (Section~\ref{ss:anti-evasion})
and what limitations our work currently has (Section~\ref{ss:extension}).

\subsection{Mitigating Evasions}
\label{ss:anti-evasion}

One way to mitigate dynamic evasions is to adapt general anti-evasion techniques from other domains to the problem of analyzing PDF documents.
Several recent papers propose to load a potentially malicious file in environments targeted at revealing the malicious behavior of the file.
For example, FuzzDroid~\cite{rasthofer2017making}, IntelliDroid~\cite{wong2016intellidroid}, and SmartDroid~\cite{zheng2012smartdroid} try to cause a potentially malicious Android app to reach ``sensitive'' API calls that would reveal malicious behavior, such as sending an SMS to a premium number.
Adapting this technique to PDF scanners requires identifying sensitive APIs in PDFs.
For known exploits, such APIs may be known, e.g., it is known that the ``Toolbutton'' exploit relies on calling the \code{app.addToolButton} API.
Finding sensitive APIs for previously unknown exploits remains an open research problem.
A related technique to cope with dynamic evasions is to explore multiple execution paths for branch decisions that depend on the environment in which a file is executed.
Rozzle~\cite{rozzle-de-cloaking-internet-malware-2} proposes this idea for client-side JavaScript code.
Adapting their approach is a promising direction for mitigating the environment-related dynamic evasions.

To deal with UI-related evasions, dynamic scanners could adapt ideas used in PuppetDroid~\cite{gianazza2014puppetdroid} and PyTrigger~\cite{fleck2013pytrigger}.
These approaches record an interaction trace from a human and play it back when loading the file under analysis to get through possible checks that guard the attack.
One of the dynamic scanners studied in this work, Cuckoo, mitigates evasions using a simpler form of this idea: The scanner arbitrarily moves the mouse to simulate a human user~\cite{cuckoo_mouse_movement}.
However, this mitigation technique is unlikely to work for evasions that require a more complicated user interaction, such as a ``captcha''.

To identify files that behave differently in an analysis environment, some techniques compare the execution behavior of the file in several different environments, e.g. virtual and physical~\cite{balzarotti2010efficient}.
Another kind of anti-evasion technique is to hide any difference between a virtual and a physical execution environment to fool the evasive malware~\cite{shi2018handling}.
Finally, to deal with the large number of possible evasions and combinations of evasions, training machine learning models to distinguish benign from malicious files seems to be a worthwhile direction~\cite{smutz2012malicious,vsrndic2013detection,laskov2011static,corona2014lux0r}.
The main challenge for effectively training machine learning models is to obtain a sufficiently large set of labeled data.
Our framework could serve as a generator of malicious training files that use different evasions and combinations of evasions.

The high recall of SAFE-PDF~\cite{2018arXiv181012490J}, which is based on abstract interpretation of JavaScript code embedded in PDFs, shows that conservative program analysis may provide an effective way of detecting malicious behavior despite evasions. The downside of any conservative program analysis are spurious warnings, which the relatively high false positive ratio of SAFE-PDF confirms.

\subsection{Choice of Scanners}
\label{ss:extension}

We focus on in-production, commercial security scanners because they represent the current state-of-the-practice, and recent academic scanners because they represent the state-of-the-art.
The studied scanners contain more static than dynamic scanners because static scanners currently dominate the market.
For example, the VirusTotal service aggregates more than 60 static scanners at the time of writing this paper~\cite{vt_engines_count}, whereas we could find only ten commercial dynamic scanners, out of which three consented to participate in this research.

Our work should not be understood as a comparison of different scanners, but rather as a comparison of each scanner's effectiveness before and after adding evasions.
The version of the scanners used in online aggregation services, which we use for the studied static scanners, may differ from the full-fledged scanners, because vendors may optimize the response time for an online service~\cite{pitfall}.

%% file: related.tex
\section{Related Work}
\label{ss: related work}

\textbf{Studying Evasions}.
Previous work has studied to what extent evasions help in circumventing scanners for malware types other than PDF.
These studies consider Android~\cite{zheng2012adam,faruki2014evaluation,rastogi2013droidchameleon,fedler2013effectiveness,petsas2014rage,canfora2015obfuscation}, Windows executables~\cite{christodorescu2004testing,moser2007limits}, and JavaScript~\cite{Xu2012a}.
Their measurements focus on reporting for each evasion whether the scanners could still detect a malicious file.
Our work differs both in the methodology and in its application.
Methodologically, our study goes beyond a single binary measure and answers additional questions, such as the added effectiveness of combined evasions and the dependence of evasions on other malware components, e.g., the payload.
Regarding the application, this work is the first to provide an in-depth study of the effectiveness of evasions for PDF-based malware.

\textbf{Analysis of PDF Malware}.
Non-executable document formats, such as PDF, have become one of the main vectors for delivering malware to victims~\cite{hardy2014targeted}.
To detect PDF documents that contain malicious JavaScript code, combinations of static and dynamic analysis of the embedded JavaScript code search for suspicious operations that rarely occur in benign documents~\cite{liu2014detecting,schmitt2012pdf,tzermias2011combining,lu2013obfuscation}.
Another line of work statically extracts features of documents, e.g., based on a document's metadata and structure~\cite{smutz2012malicious,maiorca2012pattern,vsrndic2013detection} or based on embedded JavaScript~\cite{laskov2011static,corona2014lux0r}, and then trains a machine learning model to identify malicious documents.
Nissim et al.\ survey these and other techniques~\cite{nissim2015detection}.
Beyond PDFs, the problem of malicious documents extends to other document formats~\cite{nissim2017aldocx}. 
A recurring problem for all document scanners is how to evaluate them, particularly, in the presence of evasions.
Chameleon provides a generic mechanism to create malicious documents beyond well-studied sets of documents, such as the Contagio malware dump\footnote{http://contagiodump.blogspot.com/}.

\textbf{JavaScript Analysis}.
Malicious PDF documents contain malicious JavaScript code.
Identifying such code has been actively researched for client-side web applications, by analyzing potential malware samples in a sandbox~\cite{willems2007toward}, through learning-based anomaly detection~\cite{cova2010detection}, by classifying abstract syntax trees~\cite{curtsinger2011zozzle}, or by searching for malicious sites with specially crafted search engine queries~\cite{invernizzi2012evilseed}.
A recent survey discusses various other security-related analyses of JavaScript code~\cite{jsSurvey2017}.
All these approaches focus on JavaScript code in web applications, which differs from JavaScript code embedded in PDF documents.

\textbf{From Logic Bombs to Modern Evasions}.
Attempts to fool detectors of malicious software are probably as old as malicious software itself.
Earlier approaches use logic bombs, where an attack is initiated upon occurrence of an external event~\cite{greenberg1998mobile, avivzienis2004dependability}.
To counter malware scanners that execute a potentially malicious file in a virtualized environment, anti-virtualization techniques have been proposed~\cite{raffetseder2007detecting}.
Chen et al.~\cite{chen2008towards} provide a taxonomy of malware evasion techniques with a focus on anti-virtualization and anti-debugging behavior.
Some of the evasions studied in this paper can be used to detect a virtualized environment, while others, e.g., the UI evasions, can also detect scanners running on a physical machine.
Transparent scanners try to mimic a real execution platform, i.e., without any traces of virtualization or specific fingerprints~\cite{kirat2014barecloud}, but even those can be evaded via evasion techniques that check the system's past use, e.g., via the Windows registry size or the total number of browser cookies~\cite{miramirkhani2017spotless}.
Several survey articles discuss other evasion techniques~\cite{corona2013adversarial,bulazel2017survey}, including code transformation techniques similar to our obfuscation evasions~\cite{you2010malware}.

\textbf{Evasions in Document-based Attacks}.
We envision future work to extend our framework with additional evasions, e.g., PDF parser confusion attacks~\cite{carmony2016extract}.
Other recent evasion techniques fool machine learning-based scanners, for instance by slightly modifying a benign document~\cite{Maiorca2013} or by stochastically modifying a malicious document~\cite{xu2016automatically,Dang2017}.
Knowing that attackers might conceal malicious behavior through evasions, Zhang et al.~\cite{zhang2016adversarial} propose an approach to improve machine learning-based scanners through adversary-aware feature selection.
Finally, there are two previous papers that systematically study the effectiveness of evasions.
Biggio et al.~\cite{biggio2013evasion} study to what extent learning-based malware classifiers can be fooled by evasions.
In contrast, we do not make any assumptions about the studied scanners and (probably) include both learning-based and not learning-based scanners.
Laskov et al.~\cite{laskov2014practical} focus on a single scanner (PDFRate), whereas our study involves \nbAnalyzers{} scanners.

%% file: conclusion.tex
\section{Conclusion}
\label{ss: conclusion}

% summary of approach
This paper presents a methodology to evaluate the effectiveness of evasions and its application to studying PDF malware scanners.
Our implementation of the methodology, the Chameleon framework, automatically generates and enriches malicious documents with one or multiple evasions.
We use these documents for an in-depth study of \nbAnalyzers{} PDF scanners and how they are affected by evasions.
More broadly, our methodology can also be used for studying evasions of other malware types, e.g., malicious executables.

% main take-aways
The overall result of our study is cause for concern.
We show that the studied evasions are surprisingly effective in fooling state-of-the-art scanners.
In particular by combining evasions, attackers can bypass modern defenses in both static and dynamic scanners.
Moreover, we find huge variations across scanners, enabling targeted attacks based on evasions picked specifically for a targeted scanner.
All these findings are a call to arms for future work on anti-evasion techniques.

Our work will support future efforts toward improving malware scanners in several ways.
First, the results of our study help security vendors to better understand their vulnerability to specific evasions and to focus their attention on mitigating the most effective evasions.
Second, we are releasing the corpus of malicious, evasive documents generated by Chameleon as a ready-to-use benchmark.
We are in contact with several developers of PDF scanners, and some of them, e.g., SploitGuard and SAFE-PDF, have already used our benchmark to test and improve their security scanners.
Finally, the Chameleon framework provides a basis for expanding the set of benchmarks by incorporating future evasions, exploits, and payloads.

%% file: academic_analyzers.tex
\section{Details on PDF Scanners}

\subsection{Setup of Academic and Open-Source Scanners}
\label{sec:scanner setup}

The following describes the academic and open-source scanners, i.e.,
Slayer, SAFE-PDF, PDF Scrutinizer, and Cuckoo Sandbox,
and how we set them up for our study.
In addition, we briefly go through the internals of SploitGuard, which, even though a commercial scanner, is based on an  academic work~\cite{payer2015fine}.

Slayer~\cite{maiorca2012pattern}, also known as PDF Malware Slayer or PDFMS, is a machine learning-based static scanner. It predicts whether a document is malicious based on the document's internal structure.
We train Slayer with a set of malicious and benign PDF files that are obtained from Mila Parkour, the owner of Contagiodump\footnote{http://contagiodump.blogspot.com/}, a public malware repository.
The malicious sets comprises more than 11,000 files, which are labeled as 'MALWARE\_PDF\_CVEsorted\_173\_files' and 'MALWARE\_PDF\_PRE\_04-2011\_10982\_files'.
The benign set comprises 9,000 files labeled as 'CLEAN\_PDF\_9000\_files'.

SAFE-PDF statically reasons about a file based on abstract interpretation. 
It is designed to cope with malicious PDF documents that incorporate evasions.
By abstract interpretation, it over-approximates the run-time behavior of a document to examine all its possible execution paths.

PDF Scrutinizer~\cite{schmitt2012pdf} extracts all JavaScript code snippets from a PDF document, executes the code in Mozilla Rhino, and uses libemu to find and analyze the payload.
The tool combines both static and dynamic analysis techniques, but as more weight is put on the dynamic part, we classify it as a dynamic scanner.

Cuckoo Sandbox~\cite{guarnieri2013cuckoo} (in short Cuckoo) scores each analyzed sample on a scale of 0 (benign) to 10 (certainly malicious).
To map this score into a binary score (malicious or not), which is necessary to compare Cuckoo with other scanners, we set a threshold on the score reported by Cuckoo.
To this end, we scan documents with the bare exploits, i.e., without any evasion, with Cuckoo and take the minimum score among them, 3.0, as the maliciousness threshold.
We consider any score greater than or equal to this threshold as a ``malicious'' classification, and any score smaller than the threshold as ``benign''.
We evaluate Cuckoo out-of-the-box with no additional extensions installed, except Cuckoo Signatures\footnote{https://github.com/cuckoosandbox/community/}, which are community rules that assign score to observed behaviors (e.g., dropping executable files).
We configure Cuckoo's guest machine with two processors, 2 GB of memory, and one virtual monitor having 720p resolution (important for ``mons'' and ``resol'' evasions).
The guest machine runs Windows~7, 64-bit, and has Adobe Reader 9.0, the version vulnerable to our exploits, installed.

SploitGuard is a dynamic scanner based on Lockdown~\cite{payer2015fine} that detects the exploitation of vulnerabilities by enforcing different policies during the execution of a program, for example Adobe Reader. By design, SploitGuard does not need to be trained, and the result for a given document is a binary decision ``malicious'' or ``benign''.

\subsection{Other Considered Analyzers}
\label{sec:other analyzers}

We considered several other academic scanners but could not include all of them, because some are either not available or have some issues in our local setup. The following is a list of scanners that we considered but unfortunately could not include in our study.
\begin{itemize}
    \item PDFRate~\cite{smutz2012malicious}: A learning-based PDF scanner that decides whether a document is malicious based on its structure. PDFRate's online service\footnote{https://csmutz.com/pdfrate/} was not available at the time of writing this paper.
    \item MDScan~\cite{tzermias2011combining}: Combining static and dynamic analysis, MDScan specifically targets PDF files. The dynamic analysis is via extracting JavaScript snippets and running them on an emulator. The source code for MDScan is not available in the public domain.
    \item Lux0r~\cite{corona2014lux0r}: A machine learning approach aimed at detecting malicious JavaScript code in general, but  evaluated with malicious JavaScript-bearing PDF documents. By tapping into the JavaScript interpreter, Lux0r anticipates a malicious behavior based on the API usage. Lux0r is not publicly available.
    \item MPScan~\cite{lu2013obfuscation}: MPScan extracts and de-obfuscates the JavaScript code on the fly by hooking into Adobe Reader, and then statically analyzes it to detect a malicious behavior. MPScan's source code is not publicly available.
    \item ShellOS~\cite{snow2011shellos}: Even-though designed mainly for executable files, ShellOS can find the payload in a malicious PDF document and analyze it too. However, the tool is not publicly available.
    \item The tool by Carmony et al.~\cite{carmony2016extract}: To improve the detection accuracy, Carmony et al. improve extraction of the JavaScript snippets of a PDF document. The tool then uses PJScan~\cite{laskov2011static} for classification of the extracted snippets. The tool is not publicly available.
    \item CWXDetector~\cite{willems2012using}: By disabling data execution prevention (DEP), CWXDetector monitors the execution of code from non-executable pages (former exploits usually tried to execute code from the heap, which is non-executable). Once such a write happens, the page fault handler is invoked and the page's content is dumped for further analysis. CWXDetector can be used for several file types such as executable and PDF files. However, it is not available online.
    \item Tool by Liu et al.~\cite{liu2014detecting}: By statically instrumenting a document to insert context-monitoring code, the instrumented document's behavior is observed at run-time. The tool is not publicly available.
    \item PJScan~\cite{laskov2011static}: A static scanner that uses machine learning to detect malicious files. We tried to train PJScan\footnote{https://sourceforge.net/p/pjscan/home/Home/} with the same sets of files used for training Slayer, but unfortunately, it did not find any JavaScript code (even though most of the files contain embedded JavaScript). Therefore it could not be trained and evaluated in our setup.
    \item PlatPal~\cite{xu2017platpal}: Runs a PDF document in Adobe Reader and track its behavior on Windows and macOS. Based on the discrepancies in the execution traces (e.g. the amount of dynamically allocated memory), it marks the document as either malicious or benign. We hit compilation errors while trying to build the tool\footnote{https://github.com/sslab-gatech/platpal} locally and unfortunately the documentation did not help us to resolve them.
\end{itemize}

%% file: ms.bbl

\begin{thebibliography}{80}

%%% the Web address.

\ifx \showCODEN    \undefined \def \showCODEN     #1{\unskip}     \fi
\ifx \showDOI      \undefined \def \showDOI       #1{#1}\fi
\ifx \showISBNx    \undefined \def \showISBNx     #1{\unskip}     \fi
\ifx \showISBNxiii \undefined \def \showISBNxiii  #1{\unskip}     \fi
\ifx \showISSN     \undefined \def \showISSN      #1{\unskip}     \fi
\ifx \showLCCN     \undefined \def \showLCCN      #1{\unskip}     \fi
\ifx \shownote     \undefined \def \shownote      #1{#1}          \fi
\ifx \showarticletitle \undefined \def \showarticletitle #1{#1}   \fi
\ifx \showURL      \undefined \def \showURL       {\relax}        \fi
% invisible to TeX
\providecommand\bibfield[2]{#2}
\providecommand\bibinfo[2]{#2}
\providecommand\natexlab[1]{#1}
\providecommand\showeprint[2][]{arXiv:#2}

\bibitem[\protect\citeauthoryear{??}{joe}{2014}]%
        {joe_anti_evasion}
 \bibinfo{year}{2014}\natexlab{}.
\newblock \bibinfo{title}{New Sandbox Evasion Tricks spot with Joe Sandbox
  10.5}.
\newblock
  \bibinfo{howpublished}{\url{https://www.joesecurity.org/blog/4552202539646803061}}.
\newblock
\newblock
\shownote{Accessed Dec. 2017.}


\bibitem[\protect\citeauthoryear{??}{vt_}{2017}]%
        {vt_engines_count}
 \bibinfo{year}{2017}\natexlab{}.
\newblock \bibinfo{title}{Antivirus products}.
\newblock
  \bibinfo{howpublished}{\url{https://www.virustotal.com/en/about/credits/}}.
\newblock
\newblock
\shownote{Accessed Dec. 2017.}


\bibitem[\protect\citeauthoryear{??}{pdf}{2018}]%
        {pdf_cve_statistics}
 \bibinfo{year}{2018}\natexlab{}.
\newblock \bibinfo{title}{Adobe Acrobat Reader Vulnerability Statistics}.
\newblock
  \bibinfo{howpublished}{\url{https://www.cvedetails.com/product/497/Adobe-Acrobat-Reader.html?vendor\_id=53}}.
\newblock
\newblock
\shownote{Accessed May 2018.}


\bibitem[\protect\citeauthoryear{Andreasen, Gong, M{\o}ller, Pradel, Selakovic,
  Sen, and alexandru Staicu}{Andreasen et~al\mbox{.}}{2017}]%
        {jsSurvey2017}
\bibfield{author}{\bibinfo{person}{Esben Andreasen}, \bibinfo{person}{Liang
  Gong}, \bibinfo{person}{Anders M{\o}ller}, \bibinfo{person}{Michael Pradel},
  \bibinfo{person}{Marija Selakovic}, \bibinfo{person}{Koushik Sen}, {and}
  \bibinfo{person}{Cristian alexandru Staicu}.}
  \bibinfo{year}{2017}\natexlab{}.
\newblock \showarticletitle{A Survey of Dynamic Analysis and Test Generation
  for JavaScript}.
\newblock \bibinfo{journal}{\emph{Comput. Surveys}} (\bibinfo{year}{2017}).
\newblock


\bibitem[\protect\citeauthoryear{Avi{\v{z}}ienis, Laprie, and
  Randell}{Avi{\v{z}}ienis et~al\mbox{.}}{2004}]%
        {avivzienis2004dependability}
\bibfield{author}{\bibinfo{person}{Algirdas Avi{\v{z}}ienis},
  \bibinfo{person}{Jean-Claude Laprie}, {and} \bibinfo{person}{Brian Randell}.}
  \bibinfo{year}{2004}\natexlab{}.
\newblock \showarticletitle{Dependability and its threats: a taxonomy}.
\newblock In \bibinfo{booktitle}{\emph{Building the Information Society}}.
  \bibinfo{publisher}{Springer}, \bibinfo{pages}{91--120}.
\newblock


\bibitem[\protect\citeauthoryear{Balzarotti, Cova, Karlberger, Kirda, Kruegel,
  and Vigna}{Balzarotti et~al\mbox{.}}{2010}]%
        {balzarotti2010efficient}
\bibfield{author}{\bibinfo{person}{Davide Balzarotti}, \bibinfo{person}{Marco
  Cova}, \bibinfo{person}{Christoph Karlberger}, \bibinfo{person}{Engin Kirda},
  \bibinfo{person}{Christopher Kruegel}, {and} \bibinfo{person}{Giovanni
  Vigna}.} \bibinfo{year}{2010}\natexlab{}.
\newblock \showarticletitle{Efficient Detection of Split Personalities in
  Malware}. In \bibinfo{booktitle}{\emph{NDSS}}.
\newblock


\bibitem[\protect\citeauthoryear{Besler, Willems, and Hund}{Besler
  et~al\mbox{.}}{2017}]%
        {timing_evasion}
\bibfield{author}{\bibinfo{person}{Frederic Besler}, \bibinfo{person}{Carsten
  Willems}, {and} \bibinfo{person}{Ralf Hund}.}
  \bibinfo{year}{2017}\natexlab{}.
\newblock \bibinfo{title}{Countering Innovative Sandbox Evasion Techniques Used
  by Malware}.
\newblock \bibinfo{howpublished}{29th Annual FIRST Conference}.
\newblock


\bibitem[\protect\citeauthoryear{Biggio, Corona, Maiorca, Nelson,
  {\v{S}}rndi{\'c}, Laskov, Giacinto, and Roli}{Biggio et~al\mbox{.}}{2013}]%
        {biggio2013evasion}
\bibfield{author}{\bibinfo{person}{Battista Biggio}, \bibinfo{person}{Igino
  Corona}, \bibinfo{person}{Davide Maiorca}, \bibinfo{person}{Blaine Nelson},
  \bibinfo{person}{Nedim {\v{S}}rndi{\'c}}, \bibinfo{person}{Pavel Laskov},
  \bibinfo{person}{Giorgio Giacinto}, {and} \bibinfo{person}{Fabio Roli}.}
  \bibinfo{year}{2013}\natexlab{}.
\newblock \showarticletitle{Evasion attacks against machine learning at test
  time}. In \bibinfo{booktitle}{\emph{Joint European conference on machine
  learning and knowledge discovery in databases}}. Springer,
  \bibinfo{pages}{387--402}.
\newblock


\bibitem[\protect\citeauthoryear{Bulazel and Yener}{Bulazel and Yener}{2017}]%
        {bulazel2017survey}
\bibfield{author}{\bibinfo{person}{Alexei Bulazel} {and}
  \bibinfo{person}{B{\"u}lent Yener}.} \bibinfo{year}{2017}\natexlab{}.
\newblock \showarticletitle{A Survey On Automated Dynamic Malware Analysis
  Evasion and Counter-Evasion: PC, Mobile, and Web}. In
  \bibinfo{booktitle}{\emph{Proceedings of the 1st Reversing and
  Offensive-oriented Trends Symposium}}. ACM.
\newblock


\bibitem[\protect\citeauthoryear{Canfora, Di~Sorbo, Mercaldo, and
  Visaggio}{Canfora et~al\mbox{.}}{2015}]%
        {canfora2015obfuscation}
\bibfield{author}{\bibinfo{person}{Gerardo Canfora}, \bibinfo{person}{Andrea
  Di~Sorbo}, \bibinfo{person}{Francesco Mercaldo}, {and}
  \bibinfo{person}{Corrado~Aaron Visaggio}.} \bibinfo{year}{2015}\natexlab{}.
\newblock \showarticletitle{Obfuscation techniques against signature-based
  detection: a case study}. In \bibinfo{booktitle}{\emph{Mobile Systems
  Technologies Workshop (MST), 2015}}. IEEE, \bibinfo{pages}{21--26}.
\newblock


\bibitem[\protect\citeauthoryear{Cao, Li, and Wijmans}{Cao
  et~al\mbox{.}}{2017}]%
        {gpufingerprinting}
\bibfield{author}{\bibinfo{person}{Yinzhi Cao}, \bibinfo{person}{Song Li},
  {and} \bibinfo{person}{Erik Wijmans}.} \bibinfo{year}{2017}\natexlab{}.
\newblock \showarticletitle{(Cross-)\-Browser Fingerprinting via OS and
  Hardware Level Features}.
\newblock \bibinfo{journal}{\emph{Network and Distributed System Security
  Symposium}} (\bibinfo{date}{Feb.} \bibinfo{year}{2017}).
\newblock
\urldef\tempurl%
\url{https://doi.org/10.14722/ndss.2017.23152}
\showDOI{\tempurl}


\bibitem[\protect\citeauthoryear{Carmony, Hu, Yin, Bhaskar, and Zhang}{Carmony
  et~al\mbox{.}}{2016}]%
        {carmony2016extract}
\bibfield{author}{\bibinfo{person}{Curtis Carmony}, \bibinfo{person}{Xunchao
  Hu}, \bibinfo{person}{Heng Yin}, \bibinfo{person}{Abhishek~Vasisht Bhaskar},
  {and} \bibinfo{person}{Mu Zhang}.} \bibinfo{year}{2016}\natexlab{}.
\newblock \showarticletitle{Extract Me If You Can: Abusing PDF Parsers in
  Malware Detectors.}. In \bibinfo{booktitle}{\emph{NDSS}}.
\newblock


\bibitem[\protect\citeauthoryear{Chen, Andersen, Mao, Bailey, and Nazario}{Chen
  et~al\mbox{.}}{2008}]%
        {chen2008towards}
\bibfield{author}{\bibinfo{person}{Xu Chen}, \bibinfo{person}{Jon Andersen},
  \bibinfo{person}{Z~Morley Mao}, \bibinfo{person}{Michael Bailey}, {and}
  \bibinfo{person}{Jose Nazario}.} \bibinfo{year}{2008}\natexlab{}.
\newblock \showarticletitle{Towards an understanding of anti-virtualization and
  anti-debugging behavior in modern malware}. In
  \bibinfo{booktitle}{\emph{Dependable Systems and Networks With FTCS and DCC,
  2008. DSN 2008. IEEE International Conference on}}. IEEE,
  \bibinfo{pages}{177--186}.
\newblock


\bibitem[\protect\citeauthoryear{Christodorescu and Jha}{Christodorescu and
  Jha}{2004}]%
        {christodorescu2004testing}
\bibfield{author}{\bibinfo{person}{Mihai Christodorescu} {and}
  \bibinfo{person}{Somesh Jha}.} \bibinfo{year}{2004}\natexlab{}.
\newblock \showarticletitle{Testing malware detectors}.
\newblock \bibinfo{journal}{\emph{ACM SIGSOFT Software Engineering Notes}}
  \bibinfo{volume}{29}, \bibinfo{number}{4} (\bibinfo{year}{2004}),
  \bibinfo{pages}{34--44}.
\newblock


\bibitem[\protect\citeauthoryear{Corona, Giacinto, and Roli}{Corona
  et~al\mbox{.}}{2013}]%
        {corona2013adversarial}
\bibfield{author}{\bibinfo{person}{Igino Corona}, \bibinfo{person}{Giorgio
  Giacinto}, {and} \bibinfo{person}{Fabio Roli}.}
  \bibinfo{year}{2013}\natexlab{}.
\newblock \showarticletitle{Adversarial attacks against intrusion detection
  systems: Taxonomy, solutions and open issues}.
\newblock \bibinfo{journal}{\emph{Information Sciences}}  \bibinfo{volume}{239}
  (\bibinfo{year}{2013}), \bibinfo{pages}{201--225}.
\newblock


\bibitem[\protect\citeauthoryear{Corona, Maiorca, Ariu, and Giacinto}{Corona
  et~al\mbox{.}}{2014}]%
        {corona2014lux0r}
\bibfield{author}{\bibinfo{person}{Igino Corona}, \bibinfo{person}{Davide
  Maiorca}, \bibinfo{person}{Davide Ariu}, {and} \bibinfo{person}{Giorgio
  Giacinto}.} \bibinfo{year}{2014}\natexlab{}.
\newblock \showarticletitle{Lux0r: Detection of malicious pdf-embedded
  javascript code through discriminant analysis of api references}. In
  \bibinfo{booktitle}{\emph{Proceedings of the 2014 Workshop on Artificial
  Intelligent and Security Workshop}}. ACM, \bibinfo{pages}{47--57}.
\newblock


\bibitem[\protect\citeauthoryear{Cova}{Cova}{2016}]%
        {timebased}
\bibfield{author}{\bibinfo{person}{Marco Cova}.}
  \bibinfo{year}{2016}\natexlab{}.
\newblock \bibinfo{title}{Evasive JScript}.
\newblock
  \bibinfo{howpublished}{\url{https://www.lastline.com/labsblog/evasive-jscript/}}.
\newblock
\newblock
\shownote{Accessed Jul. 2017.}


\bibitem[\protect\citeauthoryear{Cova, Kruegel, and Vigna}{Cova
  et~al\mbox{.}}{2010}]%
        {cova2010detection}
\bibfield{author}{\bibinfo{person}{Marco Cova}, \bibinfo{person}{Christopher
  Kruegel}, {and} \bibinfo{person}{Giovanni Vigna}.}
  \bibinfo{year}{2010}\natexlab{}.
\newblock \showarticletitle{Detection and analysis of drive-by-download attacks
  and malicious JavaScript code}. In \bibinfo{booktitle}{\emph{Proceedings of
  the 19th international conference on World wide web}}. ACM,
  \bibinfo{pages}{281--290}.
\newblock


\bibitem[\protect\citeauthoryear{Curtsinger, Livshits, Zorn, and
  Seifert}{Curtsinger et~al\mbox{.}}{2011}]%
        {curtsinger2011zozzle}
\bibfield{author}{\bibinfo{person}{Charlie Curtsinger},
  \bibinfo{person}{Benjamin Livshits}, \bibinfo{person}{Benjamin~G Zorn}, {and}
  \bibinfo{person}{Christian Seifert}.} \bibinfo{year}{2011}\natexlab{}.
\newblock \showarticletitle{ZOZZLE: Fast and Precise In-Browser JavaScript
  Malware Detection}. In \bibinfo{booktitle}{\emph{USENIX Security Symposium}}.
  \bibinfo{pages}{33--48}.
\newblock


\bibitem[\protect\citeauthoryear{Dang, Huang, and Chang}{Dang
  et~al\mbox{.}}{2017}]%
        {Dang2017}
\bibfield{author}{\bibinfo{person}{Hung Dang}, \bibinfo{person}{Yue Huang},
  {and} \bibinfo{person}{Ee{-}Chien Chang}.} \bibinfo{year}{2017}\natexlab{}.
\newblock \showarticletitle{Evading Classifiers by Morphing in the Dark}. In
  \bibinfo{booktitle}{\emph{Proceedings of the 2017 {ACM} {SIGSAC} Conference
  on Computer and Communications Security, {CCS} 2017, Dallas, TX, USA, October
  30 - November 03, 2017}}. \bibinfo{pages}{119--133}.
\newblock


\bibitem[\protect\citeauthoryear{Delugr\'e}{Delugr\'e}{2010}]%
        {more_related_static_obfus}
\bibfield{author}{\bibinfo{person}{Guillaume Delugr\'e}.}
  \bibinfo{year}{2010}\natexlab{}.
\newblock \bibinfo{title}{An approach to PDF shielding}.
\newblock
  \bibinfo{howpublished}{\url{http://esec-lab.sogeti.com/posts/2010/09/01/an-approach-to-pdf-shielding.html}}.
\newblock
\newblock
\shownote{Accessed May 2018.}


\bibitem[\protect\citeauthoryear{Esparza}{Esparza}{2011}]%
        {related_static_obfus}
\bibfield{author}{\bibinfo{person}{Jose~Miguel Esparza}.}
  \bibinfo{year}{2011}\natexlab{}.
\newblock \bibinfo{title}{Obfuscation and (non-)\-detection of malicious PDF
  files}.
\newblock
  \bibinfo{howpublished}{\url{http://eternal-todo.com/blog/obfuscation-non-detection-malicious-pdf-files}}.
\newblock
\newblock
\shownote{Accessed Dec. 2017.}


\bibitem[\protect\citeauthoryear{Faruki, Bharmal, Laxmi, Gaur, Conti, and
  Rajarajan}{Faruki et~al\mbox{.}}{2014}]%
        {faruki2014evaluation}
\bibfield{author}{\bibinfo{person}{Parvez Faruki}, \bibinfo{person}{Ammar
  Bharmal}, \bibinfo{person}{Vijay Laxmi}, \bibinfo{person}{Manoj~Singh Gaur},
  \bibinfo{person}{Mauro Conti}, {and} \bibinfo{person}{Muttukrishnan
  Rajarajan}.} \bibinfo{year}{2014}\natexlab{}.
\newblock \showarticletitle{Evaluation of android anti-malware techniques
  against dalvik bytecode obfuscation}. In \bibinfo{booktitle}{\emph{Trust,
  Security and Privacy in Computing and Communications (TrustCom), 2014 IEEE
  13th International Conference on}}. IEEE, \bibinfo{pages}{414--421}.
\newblock


\bibitem[\protect\citeauthoryear{Fedler, Sch{\"u}tte, and Kulicke}{Fedler
  et~al\mbox{.}}{2013}]%
        {fedler2013effectiveness}
\bibfield{author}{\bibinfo{person}{Rafael Fedler}, \bibinfo{person}{Julian
  Sch{\"u}tte}, {and} \bibinfo{person}{Marcel Kulicke}.}
  \bibinfo{year}{2013}\natexlab{}.
\newblock \showarticletitle{On the effectiveness of malware protection on
  android}.
\newblock \bibinfo{journal}{\emph{Fraunhofer AISEC}}  \bibinfo{volume}{45}
  (\bibinfo{year}{2013}).
\newblock


\bibitem[\protect\citeauthoryear{Ferrie}{Ferrie}{2007}]%
        {ferrie2007attacks}
\bibfield{author}{\bibinfo{person}{Peter Ferrie}.}
  \bibinfo{year}{2007}\natexlab{}.
\newblock \showarticletitle{Attacks on more virtual machine emulators}.
\newblock \bibinfo{journal}{\emph{Symantec Technology Exchange}}
  \bibinfo{volume}{55} (\bibinfo{year}{2007}).
\newblock


\bibitem[\protect\citeauthoryear{Fleck, Tokhtabayev, Alarif, Stavrou, and
  Nykodym}{Fleck et~al\mbox{.}}{2013}]%
        {fleck2013pytrigger}
\bibfield{author}{\bibinfo{person}{Dan Fleck}, \bibinfo{person}{Arnur
  Tokhtabayev}, \bibinfo{person}{Alex Alarif}, \bibinfo{person}{Angelos
  Stavrou}, {and} \bibinfo{person}{Tomas Nykodym}.}
  \bibinfo{year}{2013}\natexlab{}.
\newblock \showarticletitle{Pytrigger: A system to trigger \& extract
  user-activated malware behavior}. In \bibinfo{booktitle}{\emph{Availability,
  Reliability and Security (ARES), 2013 Eighth International Conference on}}.
  IEEE, \bibinfo{pages}{92--101}.
\newblock


\bibitem[\protect\citeauthoryear{Gianazza, Maggi, Fattori, Cavallaro, and
  Zanero}{Gianazza et~al\mbox{.}}{2014}]%
        {gianazza2014puppetdroid}
\bibfield{author}{\bibinfo{person}{Andrea Gianazza}, \bibinfo{person}{Federico
  Maggi}, \bibinfo{person}{Aristide Fattori}, \bibinfo{person}{Lorenzo
  Cavallaro}, {and} \bibinfo{person}{Stefano Zanero}.}
  \bibinfo{year}{2014}\natexlab{}.
\newblock \showarticletitle{Puppetdroid: A user-centric ui exerciser for
  automatic dynamic analysis of similar Android applications}.
\newblock \bibinfo{journal}{\emph{arXiv preprint arXiv:1402.4826}}
  (\bibinfo{year}{2014}).
\newblock


\bibitem[\protect\citeauthoryear{Greenberg, Byington, and Harper}{Greenberg
  et~al\mbox{.}}{1998}]%
        {greenberg1998mobile}
\bibfield{author}{\bibinfo{person}{Michael~S Greenberg},
  \bibinfo{person}{Jennifer~C Byington}, {and} \bibinfo{person}{David~G
  Harper}.} \bibinfo{year}{1998}\natexlab{}.
\newblock \showarticletitle{Mobile agents and security}.
\newblock \bibinfo{journal}{\emph{IEEE Communications magazine}}
  \bibinfo{volume}{36}, \bibinfo{number}{7} (\bibinfo{year}{1998}),
  \bibinfo{pages}{76--85}.
\newblock


\bibitem[\protect\citeauthoryear{Grooten}{Grooten}{2014}]%
        {officeonrise}
\bibfield{author}{\bibinfo{person}{Martijn Grooten}.}
  \bibinfo{year}{2014}\natexlab{}.
\newblock \bibinfo{title}{Macro malware on the rise again}.
\newblock
  \bibinfo{howpublished}{\url{https://www.virusbulletin.com/blog/2014/11/macro-malware-rise-again/}}.
\newblock
\newblock
\shownote{Accessed Jul. 2017.}


\bibitem[\protect\citeauthoryear{Guarnieri}{Guarnieri}{2012}]%
        {cuckoo_mouse_movement}
\bibfield{author}{\bibinfo{person}{Claudio Guarnieri}.}
  \bibinfo{year}{2012}\natexlab{}.
\newblock \bibinfo{title}{To the end of the World!}
\newblock
  \bibinfo{howpublished}{\url{https://cuckoosandbox.org/blog/to-the-end-of-the-world}}.
\newblock
\newblock
\shownote{Accessed May 2018.}


\bibitem[\protect\citeauthoryear{Guarnieri, Schloesser, Bremer, and
  Tanasi}{Guarnieri et~al\mbox{.}}{2013}]%
        {guarnieri2013cuckoo}
\bibfield{author}{\bibinfo{person}{C Guarnieri}, \bibinfo{person}{Mark
  Schloesser}, \bibinfo{person}{J Bremer}, {and} \bibinfo{person}{A Tanasi}.}
  \bibinfo{year}{2013}\natexlab{}.
\newblock \showarticletitle{Cuckoo sandbox-open source automated malware
  analysis}.
\newblock \bibinfo{journal}{\emph{Black Hat USA}} (\bibinfo{year}{2013}).
\newblock


\bibitem[\protect\citeauthoryear{Hardy, Crete-Nishihata, Kleemola, Senft,
  Sonne, Wiseman, Gill, and Deibert}{Hardy et~al\mbox{.}}{2014}]%
        {hardy2014targeted}
\bibfield{author}{\bibinfo{person}{Seth Hardy}, \bibinfo{person}{Masashi
  Crete-Nishihata}, \bibinfo{person}{Katharine Kleemola}, \bibinfo{person}{Adam
  Senft}, \bibinfo{person}{Byron Sonne}, \bibinfo{person}{Greg Wiseman},
  \bibinfo{person}{Phillipa Gill}, {and} \bibinfo{person}{Ronald~J Deibert}.}
  \bibinfo{year}{2014}\natexlab{}.
\newblock \showarticletitle{Targeted Threat Index: Characterizing and
  Quantifying Politically-Motivated Targeted Malware}. In
  \bibinfo{booktitle}{\emph{USENIX Security Symposium}}.
  \bibinfo{pages}{527--541}.
\newblock


\bibitem[\protect\citeauthoryear{Ho, Boneh, Ballard, and Provos}{Ho
  et~al\mbox{.}}{2014}]%
        {ho2014tick}
\bibfield{author}{\bibinfo{person}{Grant Ho}, \bibinfo{person}{Dan Boneh},
  \bibinfo{person}{Lucas Ballard}, {and} \bibinfo{person}{Niels Provos}.}
  \bibinfo{year}{2014}\natexlab{}.
\newblock \showarticletitle{Tick Tock: Building Browser Red Pills from Timing
  Side Channels}. In \bibinfo{booktitle}{\emph{WOOT}}.
\newblock


\bibitem[\protect\citeauthoryear{Invernizzi, Comparetti, Benvenuti, Kruegel,
  Cova, and Vigna}{Invernizzi et~al\mbox{.}}{2012}]%
        {invernizzi2012evilseed}
\bibfield{author}{\bibinfo{person}{Luca Invernizzi},
  \bibinfo{person}{Paolo~Milani Comparetti}, \bibinfo{person}{Stefano
  Benvenuti}, \bibinfo{person}{Christopher Kruegel}, \bibinfo{person}{Marco
  Cova}, {and} \bibinfo{person}{Giovanni Vigna}.}
  \bibinfo{year}{2012}\natexlab{}.
\newblock \showarticletitle{Evilseed: A guided approach to finding malicious
  web pages}. In \bibinfo{booktitle}{\emph{Security and Privacy (SP), 2012 IEEE
  Symposium on}}. IEEE, \bibinfo{pages}{428--442}.
\newblock


\bibitem[\protect\citeauthoryear{Itabashi}{Itabashi}{2011}]%
        {pdfstaticevasion}
\bibfield{author}{\bibinfo{person}{Kazumasa Itabashi}.}
  \bibinfo{year}{2011}\natexlab{}.
\newblock \bibinfo{booktitle}{\emph{Portable Document Format Malware}}.
\newblock \bibinfo{type}{{T}echnical {R}eport}.
  \bibinfo{institution}{Symantec}.
\newblock


\bibitem[\protect\citeauthoryear{{Jordan}, {Gauthier}, {Hassanshahi}, and
  {Zhao}}{{Jordan} et~al\mbox{.}}{2018}]%
        {2018arXiv181012490J}
\bibfield{author}{\bibinfo{person}{A. {Jordan}}, \bibinfo{person}{F.
  {Gauthier}}, \bibinfo{person}{B. {Hassanshahi}}, {and} \bibinfo{person}{D.
  {Zhao}}.} \bibinfo{year}{2018}\natexlab{}.
\newblock \showarticletitle{{SAFE-PDF: Robust Detection of JavaScript PDF
  Malware Using Abstract Interpretation}}.
\newblock \bibinfo{journal}{\emph{ArXiv e-prints}} (\bibinfo{date}{Oct.}
  \bibinfo{year}{2018}).
\newblock
\showeprint[arxiv]{cs.CR/1810.12490}


\bibitem[\protect\citeauthoryear{Keragala}{Keragala}{2016}]%
        {keragala2016detecting}
\bibfield{author}{\bibinfo{person}{Dilshan Keragala}.}
  \bibinfo{year}{2016}\natexlab{}.
\newblock \showarticletitle{Detecting malware and sandbox evasion techniques}.
\newblock \bibinfo{journal}{\emph{SANS Institute InfoSec Reading Room}}
  \bibinfo{volume}{16} (\bibinfo{year}{2016}).
\newblock


\bibitem[\protect\citeauthoryear{Kirat, Vigna, and Kruegel}{Kirat
  et~al\mbox{.}}{2014}]%
        {kirat2014barecloud}
\bibfield{author}{\bibinfo{person}{Dhilung Kirat}, \bibinfo{person}{Giovanni
  Vigna}, {and} \bibinfo{person}{Christopher Kruegel}.}
  \bibinfo{year}{2014}\natexlab{}.
\newblock \showarticletitle{BareCloud: Bare-metal Analysis-based Evasive
  Malware Detection}. In \bibinfo{booktitle}{\emph{USENIX Security Symposium}}.
  \bibinfo{pages}{287--301}.
\newblock


\bibitem[\protect\citeauthoryear{Kolbitsch}{Kolbitsch}{2014}]%
        {lastline_anti_evasion}
\bibfield{author}{\bibinfo{person}{Clemens Kolbitsch}.}
  \bibinfo{year}{2014}\natexlab{}.
\newblock \bibinfo{title}{Analyzing Environment-Aware Malware}.
\newblock
  \bibinfo{howpublished}{\url{https://www.lastline.com/labsblog/analyzing-environment-aware-malware/}}.
\newblock
\newblock
\shownote{Accessed Dec. 2017.}


\bibitem[\protect\citeauthoryear{Kolbitsch, Livshits, Zorn, and
  Seifert}{Kolbitsch et~al\mbox{.}}{2012}]%
        {rozzle-de-cloaking-internet-malware-2}
\bibfield{author}{\bibinfo{person}{Clemens Kolbitsch},
  \bibinfo{person}{Benjamin Livshits}, \bibinfo{person}{Benjamin Zorn}, {and}
  \bibinfo{person}{Christian Seifert}.} \bibinfo{year}{2012}\natexlab{}.
\newblock \showarticletitle{Rozzle: De-cloaking internet malware}. In
  \bibinfo{booktitle}{\emph{Security and Privacy (SP), 2012 IEEE Symposium
  on}}. IEEE, \bibinfo{pages}{443--457}.
\newblock


\bibitem[\protect\citeauthoryear{Laskov and {\v{S}}rndi{\'c}}{Laskov and
  {\v{S}}rndi{\'c}}{2011}]%
        {laskov2011static}
\bibfield{author}{\bibinfo{person}{Pavel Laskov} {and} \bibinfo{person}{Nedim
  {\v{S}}rndi{\'c}}.} \bibinfo{year}{2011}\natexlab{}.
\newblock \showarticletitle{Static detection of malicious JavaScript-bearing
  PDF documents}. In \bibinfo{booktitle}{\emph{Proceedings of the 27th annual
  computer security applications conference}}. ACM, \bibinfo{pages}{373--382}.
\newblock


\bibitem[\protect\citeauthoryear{Latif and Shahzad}{Latif and Shahzad}{2016}]%
        {fireeye_anti_evasion}
\bibfield{author}{\bibinfo{person}{Muhammad~Hasib Latif} {and}
  \bibinfo{person}{Farrukh Shahzad}.} \bibinfo{year}{2016}\natexlab{}.
\newblock \bibinfo{title}{Increased Use of WMI for Environment Detection and
  Evasion}.
\newblock
  \bibinfo{howpublished}{\url{https://www.fireeye.com/blog/threat-research/2016/10/increased\_use\_ofwmi.html}}.
\newblock
\newblock
\shownote{Accessed Dec. 2017.}


\bibitem[\protect\citeauthoryear{Le~Blond, Gilbert, Upadhyay, Rodriguez, and
  Choffnes}{Le~Blond et~al\mbox{.}}{2017}]%
        {le2017broad}
\bibfield{author}{\bibinfo{person}{Stevens Le~Blond},
  \bibinfo{person}{C{\'e}dric Gilbert}, \bibinfo{person}{Utkarsh Upadhyay},
  \bibinfo{person}{Manuel~Gomez Rodriguez}, {and} \bibinfo{person}{David
  Choffnes}.} \bibinfo{year}{2017}\natexlab{}.
\newblock \showarticletitle{A broad view of the ecosystem of socially
  engineered exploit documents}. In \bibinfo{booktitle}{\emph{Network and
  Distributed System Security Symposium (NDSS)}}.
\newblock


\bibitem[\protect\citeauthoryear{Liu, Wang, and Stavrou}{Liu
  et~al\mbox{.}}{2014}]%
        {liu2014detecting}
\bibfield{author}{\bibinfo{person}{Daiping Liu}, \bibinfo{person}{Haining
  Wang}, {and} \bibinfo{person}{Angelos Stavrou}.}
  \bibinfo{year}{2014}\natexlab{}.
\newblock \showarticletitle{Detecting malicious javascript in pdf through
  document instrumentation}. In \bibinfo{booktitle}{\emph{Dependable Systems
  and Networks (DSN), 2014 44th Annual IEEE/IFIP International Conference on}}.
  IEEE, \bibinfo{pages}{100--111}.
\newblock


\bibitem[\protect\citeauthoryear{Lu, Zhuge, Wang, Cao, and Chen}{Lu
  et~al\mbox{.}}{2013}]%
        {lu2013obfuscation}
\bibfield{author}{\bibinfo{person}{Xun Lu}, \bibinfo{person}{Jianwei Zhuge},
  \bibinfo{person}{Ruoyu Wang}, \bibinfo{person}{Yinzhi Cao}, {and}
  \bibinfo{person}{Yan Chen}.} \bibinfo{year}{2013}\natexlab{}.
\newblock \showarticletitle{De-obfuscation and detection of malicious pdf files
  with high accuracy}. In \bibinfo{booktitle}{\emph{System Sciences (HICSS),
  2013 46th Hawaii International Conference on}}. IEEE,
  \bibinfo{pages}{4890--4899}.
\newblock


\bibitem[\protect\citeauthoryear{Maiorca, Corona, and Giacinto}{Maiorca
  et~al\mbox{.}}{2013}]%
        {Maiorca2013}
\bibfield{author}{\bibinfo{person}{Davide Maiorca}, \bibinfo{person}{Igino
  Corona}, {and} \bibinfo{person}{Giorgio Giacinto}.}
  \bibinfo{year}{2013}\natexlab{}.
\newblock \showarticletitle{Looking at the bag is not enough to find the bomb:
  an evasion of structural methods for malicious {PDF} files detection}. In
  \bibinfo{booktitle}{\emph{8th {ACM} Symposium on Information, Computer and
  Communications Security, {ASIA} {CCS} '13, Hangzhou, China - May 08 - 10,
  2013}}. \bibinfo{pages}{119--130}.
\newblock


\bibitem[\protect\citeauthoryear{Maiorca, Giacinto, and Corona}{Maiorca
  et~al\mbox{.}}{2012}]%
        {maiorca2012pattern}
\bibfield{author}{\bibinfo{person}{Davide Maiorca}, \bibinfo{person}{Giorgio
  Giacinto}, {and} \bibinfo{person}{Igino Corona}.}
  \bibinfo{year}{2012}\natexlab{}.
\newblock \showarticletitle{A pattern recognition system for malicious pdf
  files detection}. In \bibinfo{booktitle}{\emph{International Workshop on
  Machine Learning and Data Mining in Pattern Recognition}}. Springer,
  \bibinfo{pages}{510--524}.
\newblock


\bibitem[\protect\citeauthoryear{Miramirkhani, Appini, Nikiforakis, and
  Polychronakis}{Miramirkhani et~al\mbox{.}}{2017}]%
        {miramirkhani2017spotless}
\bibfield{author}{\bibinfo{person}{Najmeh Miramirkhani},
  \bibinfo{person}{Mahathi~Priya Appini}, \bibinfo{person}{Nick Nikiforakis},
  {and} \bibinfo{person}{Michalis Polychronakis}.}
  \bibinfo{year}{2017}\natexlab{}.
\newblock \showarticletitle{Spotless sandboxes: Evading malware analysis
  systems using wear-and-tear artifacts}. In \bibinfo{booktitle}{\emph{Security
  and Privacy (SP), 2017 IEEE Symposium on}}. IEEE,
  \bibinfo{pages}{1009--1024}.
\newblock


\bibitem[\protect\citeauthoryear{Moser, Kruegel, and Kirda}{Moser
  et~al\mbox{.}}{2007}]%
        {moser2007limits}
\bibfield{author}{\bibinfo{person}{Andreas Moser}, \bibinfo{person}{Christopher
  Kruegel}, {and} \bibinfo{person}{Engin Kirda}.}
  \bibinfo{year}{2007}\natexlab{}.
\newblock \showarticletitle{Limits of static analysis for malware detection}.
  In \bibinfo{booktitle}{\emph{Computer security applications conference, 2007.
  ACSAC 2007. Twenty-third annual}}. IEEE, \bibinfo{pages}{421--430}.
\newblock


\bibitem[\protect\citeauthoryear{Nissim, Cohen, and Elovici}{Nissim
  et~al\mbox{.}}{2017}]%
        {nissim2017aldocx}
\bibfield{author}{\bibinfo{person}{Nir Nissim}, \bibinfo{person}{Aviad Cohen},
  {and} \bibinfo{person}{Yuval Elovici}.} \bibinfo{year}{2017}\natexlab{}.
\newblock \showarticletitle{ALDOCX: detection of unknown malicious microsoft
  office documents using designated active learning methods based on new
  structural feature extraction methodology}.
\newblock \bibinfo{journal}{\emph{IEEE Transactions on Information Forensics
  and Security}} \bibinfo{volume}{12}, \bibinfo{number}{3}
  (\bibinfo{year}{2017}), \bibinfo{pages}{631--646}.
\newblock


\bibitem[\protect\citeauthoryear{Nissim, Cohen, Glezer, and Elovici}{Nissim
  et~al\mbox{.}}{2015}]%
        {nissim2015detection}
\bibfield{author}{\bibinfo{person}{Nir Nissim}, \bibinfo{person}{Aviad Cohen},
  \bibinfo{person}{Chanan Glezer}, {and} \bibinfo{person}{Yuval Elovici}.}
  \bibinfo{year}{2015}\natexlab{}.
\newblock \showarticletitle{Detection of malicious PDF files and directions for
  enhancements: a state-of-the art survey}.
\newblock \bibinfo{journal}{\emph{Computers \& Security}}  \bibinfo{volume}{48}
  (\bibinfo{year}{2015}), \bibinfo{pages}{246--266}.
\newblock


\bibitem[\protect\citeauthoryear{Oh}{Oh}{2018}]%
        {exploit_CVE_2018_4990}
\bibfield{author}{\bibinfo{person}{Matt Oh}.} \bibinfo{year}{2018}\natexlab{}.
\newblock \bibinfo{title}{Taking apart a double zero-day sample discovered in
  joint hunt with ESET}.
\newblock
  \bibinfo{howpublished}{\url{https://cloudblogs.microsoft.com/microsoftsecure/2018/07/02/taking-apart-a-double-zero-day-sample-discovered-in-joint-hunt-with-eset/}}.
\newblock
\newblock
\shownote{Accessed Aug 2018.}


\bibitem[\protect\citeauthoryear{\'Ov\'ari}{\'Ov\'ari}{2015}]%
        {pdf_obfus_odd_algs}
\bibfield{author}{\bibinfo{person}{D\'enes \'Ov\'ari}.}
  \bibinfo{year}{2015}\natexlab{}.
\newblock \bibinfo{title}{Script in a lossy stream}.
\newblock
  \bibinfo{howpublished}{\url{https://www.virusbulletin.com/virusbulletin/2015/03/script-lossy-stream}}.
\newblock
\newblock
\shownote{Accessed Dec. 2017.}


\bibitem[\protect\citeauthoryear{Payer, Barresi, and Gross}{Payer
  et~al\mbox{.}}{2015}]%
        {payer2015fine}
\bibfield{author}{\bibinfo{person}{Mathias Payer}, \bibinfo{person}{Antonio
  Barresi}, {and} \bibinfo{person}{Thomas~R Gross}.}
  \bibinfo{year}{2015}\natexlab{}.
\newblock \showarticletitle{Fine-grained control-flow integrity through binary
  hardening}. In \bibinfo{booktitle}{\emph{International Conference on
  Detection of Intrusions and Malware, and Vulnerability Assessment}}.
  Springer, \bibinfo{pages}{144--164}.
\newblock


\bibitem[\protect\citeauthoryear{Petsas, Voyatzis, Athanasopoulos,
  Polychronakis, and Ioannidis}{Petsas et~al\mbox{.}}{2014}]%
        {petsas2014rage}
\bibfield{author}{\bibinfo{person}{Thanasis Petsas}, \bibinfo{person}{Giannis
  Voyatzis}, \bibinfo{person}{Elias Athanasopoulos}, \bibinfo{person}{Michalis
  Polychronakis}, {and} \bibinfo{person}{Sotiris Ioannidis}.}
  \bibinfo{year}{2014}\natexlab{}.
\newblock \showarticletitle{Rage against the virtual machine: hindering dynamic
  analysis of android malware}. In \bibinfo{booktitle}{\emph{Proceedings of the
  Seventh European Workshop on System Security}}. ACM, \bibinfo{pages}{5}.
\newblock


\bibitem[\protect\citeauthoryear{Quintero}{Quintero}{2012}]%
        {pitfall}
\bibfield{author}{\bibinfo{person}{Bernardo Quintero}.}
  \bibinfo{year}{2012}\natexlab{}.
\newblock \bibinfo{title}{AV Comparative Analyses, Marketing, and VirusTotal: A
  Bad Combination}.
\newblock
  \bibinfo{howpublished}{\url{http://blog.virustotal.com/2012/08/av-comparative-analyses-marketing-and.html}}.
\newblock
\newblock
\shownote{Accessed Dec. 2017.}


\bibitem[\protect\citeauthoryear{Raffetseder, Kruegel, and Kirda}{Raffetseder
  et~al\mbox{.}}{2007}]%
        {raffetseder2007detecting}
\bibfield{author}{\bibinfo{person}{Thomas Raffetseder},
  \bibinfo{person}{Christopher Kruegel}, {and} \bibinfo{person}{Engin Kirda}.}
  \bibinfo{year}{2007}\natexlab{}.
\newblock \showarticletitle{Detecting system emulators}. In
  \bibinfo{booktitle}{\emph{International Conference on Information Security}}.
  Springer, \bibinfo{pages}{1--18}.
\newblock


\bibitem[\protect\citeauthoryear{Rasthofer, Arzt, Triller, and
  Pradel}{Rasthofer et~al\mbox{.}}{2017}]%
        {rasthofer2017making}
\bibfield{author}{\bibinfo{person}{Siegfried Rasthofer},
  \bibinfo{person}{Steven Arzt}, \bibinfo{person}{Stefan Triller}, {and}
  \bibinfo{person}{Michael Pradel}.} \bibinfo{year}{2017}\natexlab{}.
\newblock \showarticletitle{Making malory behave maliciously: Targeted fuzzing
  of Android execution environments}. In \bibinfo{booktitle}{\emph{Proceedings
  of the 39th International Conference on Software Engineering}}. IEEE Press,
  \bibinfo{pages}{300--311}.
\newblock


\bibitem[\protect\citeauthoryear{Rastogi, Chen, and Jiang}{Rastogi
  et~al\mbox{.}}{2013}]%
        {rastogi2013droidchameleon}
\bibfield{author}{\bibinfo{person}{Vaibhav Rastogi}, \bibinfo{person}{Yan
  Chen}, {and} \bibinfo{person}{Xuxian Jiang}.}
  \bibinfo{year}{2013}\natexlab{}.
\newblock \showarticletitle{Droidchameleon: evaluating android anti-malware
  against transformation attacks}. In \bibinfo{booktitle}{\emph{Proceedings of
  the 8th ACM SIGSAC symposium on Information, computer and communications
  security}}. ACM, \bibinfo{pages}{329--334}.
\newblock


\bibitem[\protect\citeauthoryear{Roccia}{Roccia}{2017}]%
        {historyofevasion}
\bibfield{author}{\bibinfo{person}{Thomas Roccia}.}
  \bibinfo{year}{2017}\natexlab{}.
\newblock \bibinfo{title}{Malware evasion techniques and trends}.
\newblock \bibinfo{howpublished}{McAfee Labs Threats Report}.
\newblock


\bibitem[\protect\citeauthoryear{Schmitt, Gassen, and Gerhards-Padilla}{Schmitt
  et~al\mbox{.}}{2012}]%
        {schmitt2012pdf}
\bibfield{author}{\bibinfo{person}{Florian Schmitt}, \bibinfo{person}{Jan
  Gassen}, {and} \bibinfo{person}{Elmar Gerhards-Padilla}.}
  \bibinfo{year}{2012}\natexlab{}.
\newblock \showarticletitle{PDF SCRUTINIZER: Detecting JavaScript-based attacks
  in PDF documents}. In \bibinfo{booktitle}{\emph{Privacy, Security and Trust
  (PST), 2012 Tenth Annual International Conference on}}. IEEE,
  \bibinfo{pages}{104--111}.
\newblock


\bibitem[\protect\citeauthoryear{Shi, Mirkovic, and Alwabel}{Shi
  et~al\mbox{.}}{2018}]%
        {shi2018handling}
\bibfield{author}{\bibinfo{person}{Hao Shi}, \bibinfo{person}{Jelena Mirkovic},
  {and} \bibinfo{person}{Abdulla Alwabel}.} \bibinfo{year}{2018}\natexlab{}.
\newblock \showarticletitle{Handling Anti-Virtual Machine Techniques in
  Malicious Software}.
\newblock \bibinfo{journal}{\emph{ACM Transactions on Privacy and Security
  (TOPS)}} \bibinfo{volume}{21}, \bibinfo{number}{1} (\bibinfo{year}{2018}).
\newblock


\bibitem[\protect\citeauthoryear{Smutz and Stavrou}{Smutz and Stavrou}{2012}]%
        {smutz2012malicious}
\bibfield{author}{\bibinfo{person}{Charles Smutz} {and}
  \bibinfo{person}{Angelos Stavrou}.} \bibinfo{year}{2012}\natexlab{}.
\newblock \showarticletitle{Malicious PDF detection using metadata and
  structural features}. In \bibinfo{booktitle}{\emph{Proceedings of the 28th
  annual computer security applications conference}}. ACM,
  \bibinfo{pages}{239--248}.
\newblock


\bibitem[\protect\citeauthoryear{Snow, Krishnan, Monrose, and Provos}{Snow
  et~al\mbox{.}}{2011}]%
        {snow2011shellos}
\bibfield{author}{\bibinfo{person}{Kevin~Z Snow}, \bibinfo{person}{Srinivas
  Krishnan}, \bibinfo{person}{Fabian Monrose}, {and} \bibinfo{person}{Niels
  Provos}.} \bibinfo{year}{2011}\natexlab{}.
\newblock \showarticletitle{SHELLOS: Enabling Fast Detection and Forensic
  Analysis of Code Injection Attacks.}. In \bibinfo{booktitle}{\emph{USENIX
  Security Symposium}}. \bibinfo{pages}{183--200}.
\newblock


\bibitem[\protect\citeauthoryear{{\v{S}}rndic and Laskov}{{\v{S}}rndic and
  Laskov}{2013}]%
        {vsrndic2013detection}
\bibfield{author}{\bibinfo{person}{Nedim {\v{S}}rndic} {and}
  \bibinfo{person}{Pavel Laskov}.} \bibinfo{year}{2013}\natexlab{}.
\newblock \showarticletitle{Detection of malicious pdf files based on
  hierarchical document structure}. In \bibinfo{booktitle}{\emph{Proceedings of
  the 20th Annual Network \& Distributed System Security Symposium}}.
  \bibinfo{pages}{1--16}.
\newblock


\bibitem[\protect\citeauthoryear{{\v{S}}rndi{\'c} and Laskov}{{\v{S}}rndi{\'c}
  and Laskov}{2014}]%
        {laskov2014practical}
\bibfield{author}{\bibinfo{person}{Nedim {\v{S}}rndi{\'c}} {and}
  \bibinfo{person}{Pavel Laskov}.} \bibinfo{year}{2014}\natexlab{}.
\newblock \showarticletitle{Practical evasion of a learning-based classifier: A
  case study}. In \bibinfo{booktitle}{\emph{Security and Privacy (SP), 2014
  IEEE Symposium on}}. IEEE, \bibinfo{pages}{197--211}.
\newblock


\bibitem[\protect\citeauthoryear{Townsend}{Townsend}{2016}]%
        {static_analyzers_share_analysis_result}
\bibfield{author}{\bibinfo{person}{Kevin Townsend}.}
  \bibinfo{year}{2016}\natexlab{}.
\newblock \bibinfo{title}{VirusTotal Policy Change Rocks Anti-Malware
  Industry}.
\newblock
  \bibinfo{howpublished}{\url{http://www.securityweek.com/virustotal-policy-change-rocks-anti-malware-industry}}.
\newblock
\newblock
\shownote{Accessed Dec. 2017.}


\bibitem[\protect\citeauthoryear{Tzermias, Sykiotakis, Polychronakis, and
  Markatos}{Tzermias et~al\mbox{.}}{2011}]%
        {tzermias2011combining}
\bibfield{author}{\bibinfo{person}{Zacharias Tzermias},
  \bibinfo{person}{Giorgos Sykiotakis}, \bibinfo{person}{Michalis
  Polychronakis}, {and} \bibinfo{person}{Evangelos~P Markatos}.}
  \bibinfo{year}{2011}\natexlab{}.
\newblock \showarticletitle{Combining static and dynamic analysis for the
  detection of malicious documents}. In \bibinfo{booktitle}{\emph{Proceedings
  of the Fourth European Workshop on System Security}}. ACM.
\newblock


\bibitem[\protect\citeauthoryear{Willems}{Willems}{2016a}]%
        {vmray_anti_evasion}
\bibfield{author}{\bibinfo{person}{Carsten Willems}.}
  \bibinfo{year}{2016}\natexlab{a}.
\newblock \bibinfo{title}{Nowhere to Hide: Analyzing Environment-Sensitive
  Malware with Rewind}.
\newblock
  \bibinfo{howpublished}{\url{https://www.vmray.com/blog/analyzing-environment-sensitive-malware/}}.
\newblock
\newblock
\shownote{Accessed Dec. 2017.}


\bibitem[\protect\citeauthoryear{Willems}{Willems}{2016b}]%
        {userinteraction}
\bibfield{author}{\bibinfo{person}{Carsten Willems}.}
  \bibinfo{year}{2016}\natexlab{b}.
\newblock \bibinfo{title}{Sandbox Evasion Techniques Part 4}.
\newblock
  \bibinfo{howpublished}{\url{https://www.vmray.com/blog/sandbox-evasion-techniques-part-4/}}.
\newblock
\newblock
\shownote{Accessed Jul. 2017.}


\bibitem[\protect\citeauthoryear{Willems, Freiling, and Holz}{Willems
  et~al\mbox{.}}{2012}]%
        {willems2012using}
\bibfield{author}{\bibinfo{person}{Carsten Willems}, \bibinfo{person}{Felix~C
  Freiling}, {and} \bibinfo{person}{Thorsten Holz}.}
  \bibinfo{year}{2012}\natexlab{}.
\newblock \showarticletitle{Using memory management to detect and extract
  illegitimate code for malware analysis}. In
  \bibinfo{booktitle}{\emph{Proceedings of the 28th Annual Computer Security
  Applications Conference}}. ACM, \bibinfo{pages}{179--188}.
\newblock


\bibitem[\protect\citeauthoryear{Willems, Holz, and Freiling}{Willems
  et~al\mbox{.}}{2007}]%
        {willems2007toward}
\bibfield{author}{\bibinfo{person}{Carsten Willems}, \bibinfo{person}{Thorsten
  Holz}, {and} \bibinfo{person}{Felix Freiling}.}
  \bibinfo{year}{2007}\natexlab{}.
\newblock \showarticletitle{Toward automated dynamic malware analysis using
  cwsandbox}.
\newblock \bibinfo{journal}{\emph{IEEE Security \& Privacy}}
  \bibinfo{volume}{5}, \bibinfo{number}{2} (\bibinfo{year}{2007}).
\newblock


\bibitem[\protect\citeauthoryear{Wong and Lie}{Wong and Lie}{2016}]%
        {wong2016intellidroid}
\bibfield{author}{\bibinfo{person}{Michelle~Y Wong} {and}
  \bibinfo{person}{David Lie}.} \bibinfo{year}{2016}\natexlab{}.
\newblock \showarticletitle{IntelliDroid: A Targeted Input Generator for the
  Dynamic Analysis of Android Malware}. In \bibinfo{booktitle}{\emph{NDSS}}.
\newblock


\bibitem[\protect\citeauthoryear{Xu and Kim}{Xu and Kim}{2017}]%
        {xu2017platpal}
\bibfield{author}{\bibinfo{person}{Meng Xu} {and} \bibinfo{person}{Taesoo
  Kim}.} \bibinfo{year}{2017}\natexlab{}.
\newblock \showarticletitle{PlatPal: Detecting Malicious Documents with
  Platform Diversity}. In \bibinfo{booktitle}{\emph{USENIX Security 2017}}.
  \bibinfo{pages}{271--287}.
\newblock


\bibitem[\protect\citeauthoryear{Xu, Qi, and Evans}{Xu et~al\mbox{.}}{2016}]%
        {xu2016automatically}
\bibfield{author}{\bibinfo{person}{Weilin Xu}, \bibinfo{person}{Yanjun Qi},
  {and} \bibinfo{person}{David Evans}.} \bibinfo{year}{2016}\natexlab{}.
\newblock \showarticletitle{Automatically evading classifiers}. In
  \bibinfo{booktitle}{\emph{Proceedings of the 2016 Network and Distributed
  Systems Symposium}}.
\newblock


\bibitem[\protect\citeauthoryear{Xu, Zhang, and Zhu}{Xu et~al\mbox{.}}{2012}]%
        {Xu2012a}
\bibfield{author}{\bibinfo{person}{Wei Xu}, \bibinfo{person}{Fangfang Zhang},
  {and} \bibinfo{person}{Sencun Zhu}.} \bibinfo{year}{2012}\natexlab{}.
\newblock \showarticletitle{The power of obfuscation techniques in malicious
  JavaScript code: {A} measurement study}. In \bibinfo{booktitle}{\emph{7th
  International Conference on Malicious and Unwanted Software, {MALWARE} 2012,
  Fajardo, PR, USA, October 16-18, 2012}}. \bibinfo{pages}{9--16}.
\newblock


\bibitem[\protect\citeauthoryear{You and Yim}{You and Yim}{2010}]%
        {you2010malware}
\bibfield{author}{\bibinfo{person}{Ilsun You} {and} \bibinfo{person}{Kangbin
  Yim}.} \bibinfo{year}{2010}\natexlab{}.
\newblock \showarticletitle{Malware obfuscation techniques: A brief survey}. In
  \bibinfo{booktitle}{\emph{Broadband, Wireless Computing, Communication and
  Applications (BWCCA), 2010 International Conference on}}. IEEE,
  \bibinfo{pages}{297--300}.
\newblock


\bibitem[\protect\citeauthoryear{Zhang, Chan, Biggio, Yeung, and Roli}{Zhang
  et~al\mbox{.}}{2016}]%
        {zhang2016adversarial}
\bibfield{author}{\bibinfo{person}{Fei Zhang}, \bibinfo{person}{Patrick~PK
  Chan}, \bibinfo{person}{Battista Biggio}, \bibinfo{person}{Daniel~S Yeung},
  {and} \bibinfo{person}{Fabio Roli}.} \bibinfo{year}{2016}\natexlab{}.
\newblock \showarticletitle{Adversarial feature selection against evasion
  attacks}.
\newblock \bibinfo{journal}{\emph{IEEE transactions on cybernetics}}
  \bibinfo{volume}{46}, \bibinfo{number}{3} (\bibinfo{year}{2016}),
  \bibinfo{pages}{766--777}.
\newblock


\bibitem[\protect\citeauthoryear{Zheng, Zhu, Dai, Gu, Gong, Han, and Zou}{Zheng
  et~al\mbox{.}}{2012b}]%
        {zheng2012smartdroid}
\bibfield{author}{\bibinfo{person}{Cong Zheng}, \bibinfo{person}{Shixiong Zhu},
  \bibinfo{person}{Shuaifu Dai}, \bibinfo{person}{Guofei Gu},
  \bibinfo{person}{Xiaorui Gong}, \bibinfo{person}{Xinhui Han}, {and}
  \bibinfo{person}{Wei Zou}.} \bibinfo{year}{2012}\natexlab{b}.
\newblock \showarticletitle{Smartdroid: an automatic system for revealing
  UI-based trigger conditions in android applications}. In
  \bibinfo{booktitle}{\emph{Proceedings of the second ACM workshop on Security
  and privacy in smartphones and mobile devices}}. ACM,
  \bibinfo{pages}{93--104}.
\newblock


\bibitem[\protect\citeauthoryear{Zheng, Lee, and Lui}{Zheng
  et~al\mbox{.}}{2012a}]%
        {zheng2012adam}
\bibfield{author}{\bibinfo{person}{Min Zheng}, \bibinfo{person}{Patrick~PC
  Lee}, {and} \bibinfo{person}{John~CS Lui}.} \bibinfo{year}{2012}\natexlab{a}.
\newblock \showarticletitle{ADAM: an automatic and extensible platform to
  stress test android anti-virus systems}. In
  \bibinfo{booktitle}{\emph{International conference on detection of intrusions
  and malware, and vulnerability assessment}}. Springer,
  \bibinfo{pages}{82--101}.
\newblock


\end{thebibliography}
